\documentclass{aastex61}

\newcommand{\um }{$\mu $m}
\newcommand{\cmq}{cm{$^{-3}$}}

\newcommand{\kms}{km~s{$^{-1}$}}

\newcommand{\Msol}{M{$_{\odot}$}}

\newcommand{\Vlsr}{{V$_{LSR}$}}

\newcommand{\Hi}{H{\sc i}}

\newcommand{\Htwo}{H{$_2$}}

\newcommand{\yr}{yr$^{-1}$}

\received{August  1, 2019}
\revised{November 10, 2019}
\submitjournal{ApJ}

\shorttitle{OMC1 dynamic ejection}
\shortauthors{Bally  et al.}

\begin{document}

\title{The Orion Protostellar Explosion and Runaway Stars Revisited: \\
Stellar Masses, Disk Retention, and an Outflow from BN.}

\correspondingauthor{John Bally}
\email{john.bally@colorado.edu}

\author[0000-0001-8135-6612]{John Bally}
\affil{ Center for Astrophysics and Space Astronomy,
       Astrophysical and Planetary Sciences Department,
       University of Colorado, UCB 389 Boulder, Colorado 80309, USA}
       
\author{Adam Ginsburg}
\affil{Department of Astronomy,
        University of Florida,
        Bryant Space Science Center,
        Stadium Road, 
        Gainesville, FL 32611}

\author{Jan Forbrich}
\affil{ Centre for Astrophysics Research, School of Physics, 
        Astronomy and Mathematics, Hertfordshire University, 
        College Lane, Hatfield, AL10 9AB, UK }

\author{Jaime Vargas-Gonz\'{a}lez }
\affil{ Centre for Astrophysics Research, School of Physics, 
        Astronomy and Mathematics, Hertfordshire University, 
        College Lane, Hatfield, AL10 9AB, UK }
        

\begin{abstract}

The proper motions of the three stars ejected from Orion's OMC1 cloud 
core are combined with the requirement  that their center of mass is
gravitationally bound to OMC1 to show that radio source I (Src I)  
is likely to have a mass around 15  \Msol\  consistent with recent
measurements.   Src I, the star with the 
smallest proper motion, is suspected to be either an AU-scale  
binary or a protostellar merger remnant produced by a dynamic interaction 
$\sim$550 years ago.   Near-infrared 2.2 \um\ images 
spanning $\sim$21 years confirm the $\sim$55 \kms\  motion of  
`source x' (Src x)  away from the site of stellar ejection and point 
of origin of the explosive OMC1 protostellar outflow.   The radial 
velocities and masses of the  Becklin-Neugebauer (BN) object and 
Src I constrain the radial  velocity of Src x to be  
$\rm V_{LSR} = -28 \pm 10$~\kms .  
Several high proper-motion radio sources near BN, including 
Zapata 11 ([ZRK2004] 11)
and a diffuse source near IRc 23,  may trace a slow bipolar 
outflow from BN.   The massive disk around Src I is likely the 
surviving  portion of a disk  that existed  prior to the stellar ejection.  
Though highly perturbed, shocked,  and re-oriented by the N-body 
interaction,  enough time has elapsed to allow the disk to relax 
with its spin axis roughly orthogonal to the proper motion.

\end{abstract}


\section{Introduction} \label{sec:intro}

The Orion OMC1 cloud core located a few tenths of a parsec behind the Orion Nebula 
contains the nearest site of on-going massive star formation \citep{ODell2001,Bally2016}.    
The  outflow emerging from OMC1 has a kinetic energy 
of 2$-$4$\times 10^{47}$ ergs today and exhibits a remarkable explosive  morphology 
\citep{Zapata2009,Bally2011,Bally2015,Bally2016,Bally2017}.  This outflow originated 
from within a few arc-seconds of the location from which three stars were ejected with 
speeds of $\sim$10 to 55 \kms\ about 550 years ago.  
The Becklin-Neugebauer (BN) object has a proper motion of $\sim$ 26 \kms\
towards the northwest,  radio source Src I moves with $\sim$10 \kms\ towards the 
south, and near-infrared source 
Src~x moves with $\sim$55 \kms\ toward the southeast 
\citep{Dzib2017,Rodriguez2017,Luhman2017}.  
The stellar ejection and the explosion in the gas appear to have been triggered by 
the same event,   likely caused by an N-body stellar interaction that led to either the 
formation of an AU-scale binary or a  stellar merger which released at least 
$10^{48}$ ergs of gravitational potential energy required to power the explosion
and the acceleration of the runaway stars 
\citep{Zapata2009, Goddi2011,  Bally2017, FariasTan2018}.   

Before 2017,  infrared source n  (Src n) was thought to be a runaway star from
OMC1 because radio emission from this object exhibited high proper motions.  
However, \citet{ Rodriguez2017} showed that radio proper motions were biased
by the ejection of a one-sided radio jet around 2006, implying that source n
may not have a measurable proper motion in the Orion reference frame.   
However, near-infrared Hubble Space Telescope (HST) images revealed 
that  \citet{Lonsdale1982} `source x' (Src x), has  a  $\sim$55 \kms\ 
proper motion away from the explosion center \citep{Luhman2017}.

SiO masers and other emission lines tracing the Src I disk show that it 
has a radial velocity  \Vlsr =5.0 \kms\  \citep{PlambeckWright2016}.  
\citet{Ginsburg2018} and  \citet{Ginsburg2019} found \Vlsr = 5.5 \kms\  
using thermal emission lines.    The OMC1 core has a radial velocity 
V$_{OMC1}$= 9.0 to 9.5 \kms .  Thus we
adopt a value of  $V_r(I)$=$-$4$\pm0.5$ \kms\   for the radial velocity of Src I
in the Orion reference frame.   Hydrogen recombination 
lines from BN show that it has a radial velocity \Vlsr =23.2$\pm$0.5 \kms ,
implying a $V_r(BN)$=+14.2$\pm 0.5$ \kms\ in the Orion 
reference frame \citep{Plambeck2013}.   
These radial velocity measurements are combined with proper motion
measurements of the three ejected stars to estimate their 
velocity vectors.

The mass of the BN object, thought to be an early B star, has been estimated to be 
in the range 10 to 15 \Msol\ based on spectroscopy, its luminosity, and the presence
of hydrogen recombination lines and free-free emission from a hypercompact HII region  
\citep{Scoville1983,Jiang2005,Plambeck2013}.   The mass of Src x 
is estimated to be roughly 3 \Msol\ based on its IR spectrum \citep{Luhman2017}.

Unfortunately, Src I is so embedded that it is only visible at sub-mm to radio
wavelengths.  Thus, its mass can't be determined from its stellar spectrum.  
However,  it is surrounded by a circumstellar disk whose rotation curve 
has been used  to estimate its mass dynamically.    Measurements  
with an angular resolution as high as 0.2\arcsec\ resulted in a masses around 
5 to 8 \Msol\  using the so-called centroid-of-channels method
\citep{Kim2008,Matthews2010,Hirota2014,PlambeckWright2016,Hirota2017,
Kim2019}.   

\citet{Ginsburg2018} used ALMA to study the disk rotation curve with
the highest angular resolution achieved thus far, $\sim$0.03\arcsec .  Three emission
lines were used;  the 232.6867 GHz H$_2$O $\rm 5_{5,0}-6_{4,3}$ transition 
whose upper state is 3461.9 K above ground and transitions at 
217.980 GHz and 232.511 GHz from (at the time of publication) an unknown carrier.    
Fitting the maximum radial velocity at each position in the nearly edge-on disk, 
\citet{Ginsburg2018} derived a mass of $\sim$15$\pm$2 \Msol\ for Src I. 
The mass measurement was verified using radiative transfer modeling 
analogous to that used by \citet{PlambeckWright2016}.  \citet{Ginsburg2019} 
show that the two unknown lines are transitions of salt, NaCl, v=2, J=17-16 and 
v=1, J=18-17, respectively, with upper states 1128 K and 626 K above ground.  

\citet{FariasTan2018} presented $\sim 10^7$ N-body scattering simulations
of the stellar dynamical ejection. 
They found that a binary-binary  interaction has the highest probability of  
leading to the observed configuration of three runaway stars, Src I, BN, and Src x.   
In their model, they consider Src I to be a compact binary.  They also explore the 
system center of mass velocity  as a function of the initial stellar masses and motions.   
They show that the mass of Src I (total mass of a binary or a single star)  
is likely to be between 5 and 25 \Msol\ with a most likely value near 14 \Msol . 

In this paper, we present archival observations which provide additional constraints
on the motions of stars in the OMC1 region.  We use the measured proper motions,
a range of estimated masses for BN and Src x,  and the requirement 
that the center of mass of the stars moves with a speed less than the gravitational 
escape speed from OMC1 to show that Src I must have a mass around 15  \Msol .
We use the measured radial velocities of BN and Src I to predict the 
radial velocity of Src x which thus far has not been measured.   

The distance to OMC1 had been determined to be D$\approx$414$\pm$7 pc 
using radio parallaxes to young stars exhibiting non-thermal emission \citep{Menten2007}
and the H$_2$O masers in the outflow  yielding D$\approx$418$\pm$6 pc \citep{Kim2008}.
More recent parallax measurements to radio continuum emitting stars in the Orion A 
molecular cloud which hosts OMC1 yield a distance of  388$\pm$5 pc \citep{Kounkel2017}.
Several shocks powered by the OMC1 outflow protrude into the Orion Nebula,
indicating that the OMC1 core must be only a few tenths of a parsec behind the 
nebula's ionization front and not far behind most of the stars measured by
\citet{Kounkel2017}.   We adopt a distance of 400 pc to the OMC1 core, the
explosive outflow and associated stars and use this value in all calculations. 

In addition to Src I and BN, several additional compact radio continuum features 
near BN exhibit  proper motions away from the explosion and stellar 
ejection center \citep{Dzib2017}.   

A 1\arcsec\ diameter radio source located  7.7\arcsec\ east of BN at 
position angle PA $\sim$ 77\arcdeg\   and  dubbed `IRc23' was found to move 
with a velocity of $\sim$39 \kms\ towards PA $\sim$9\arcdeg\  \citep{Dzib2017}.
IRc23 is actually an infrared source \citep{Shuping2004} located about 1.3\arcsec\ 
northeast of this radio feature and is associated with a bipolar disk shadow in 
images at $\lambda$ = 1.64 and 2.12 $\mu$m \citep{Bally2015}.  The source 
[ZRK] 11 \citep{Zapata2004}, referred to as Zapata 11 in the rest of this manuscript, 
is located  $\sim$2\arcsec\ west of BN.  According to  \citet{Dzib2017}, Zapata 11 
has a velocity 28 \kms\ towards  PA $\sim$308\arcdeg , similar to the motion of BN itself.

Neither of these 
slightly extended  radio sources have X-ray, infrared,  or millimeter-wave 
counterparts.  Thus, they  are unlikely to contain stars.  
These radio features are likely to be ionized
ejecta produced by the OMC1 explosion or powered by an outflow from BN itself.  
Therefore, they probably have negligible mass and will be ignored in our 
center of momentum analysis.

Figure~1 shows the central part of the OMC1 
explosion in an image obtained through a $\sim$1\% band-pass narrow-band 
filter transmitting the 2.12 $\mu$m S(1) ro-vibrational transition of \Htwo  .  
This image, taken from \citet{Bally2015} has a resolution of $\sim$0.06\arcsec\ 
and  is reproduced here to show the proper motion of Src x which was not 
known to be a runaway star in 2015.  The locations and proper motions of 
the three ejected  stars are indicated by the  cyan labels and yellow vectors, 
respectively.





\section{Observations and Analysis}

\begin{figure*}
\plotone{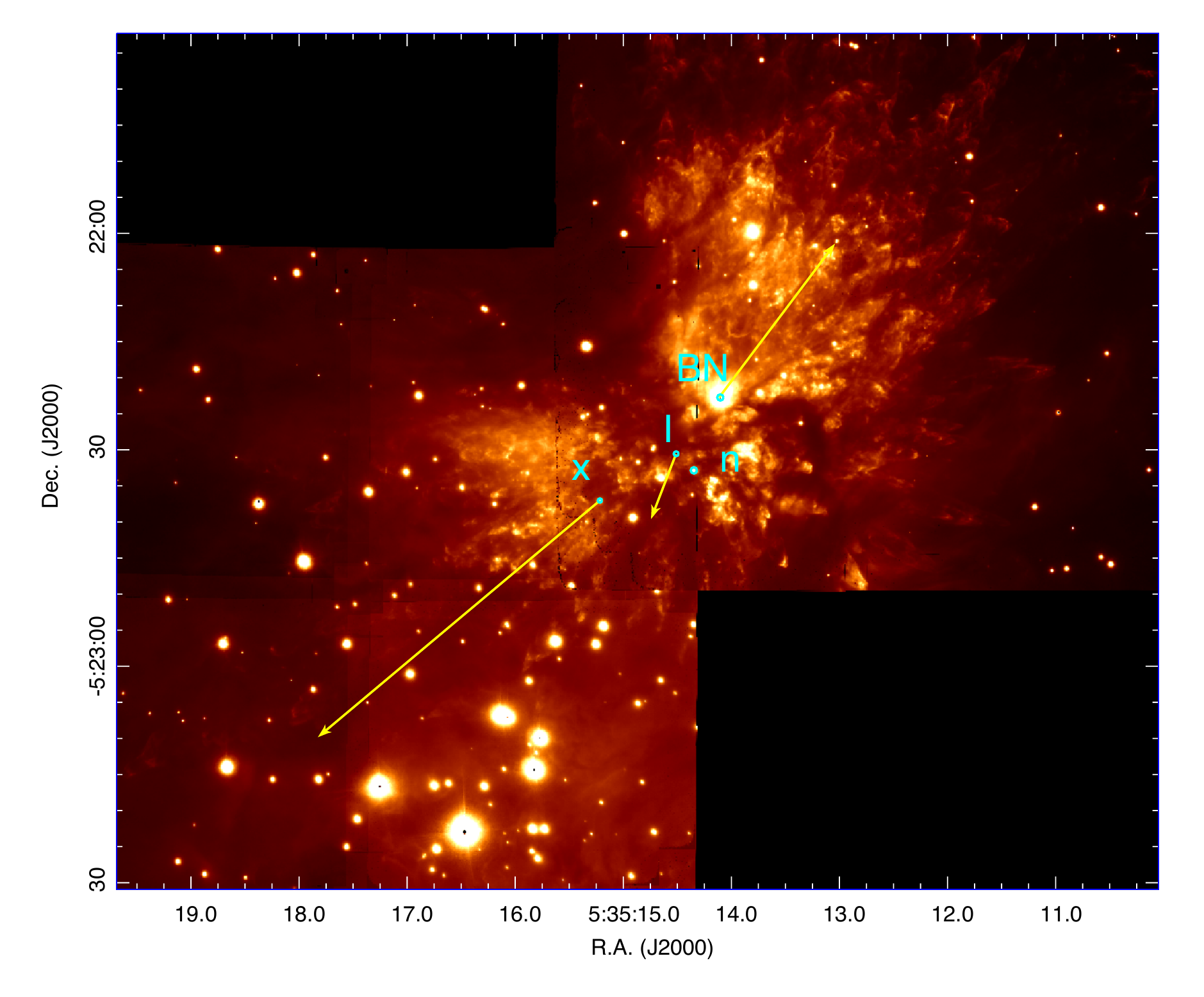}
\caption{A 2.12 $\mu$m narrow band \Htwo\ image from  Bally et al. (2015) showing  Src x, Src I, BN, and Src n.   The yellow arrows show the proper motions of the three ejected stars with lengths corresponding to the motions over the next 2,000 years.  The vectors shown are based on the motions measured by \citet{Rodriguez2017} for the BN object and Src I, and by \citet{Luhman2017} for Src x.  The Trapezium cluster of massive stars in the center of the Orion Nebula are located near the bottom of the image.
}
\label{fig1}
\end{figure*}

\begin{figure*}
\plotone{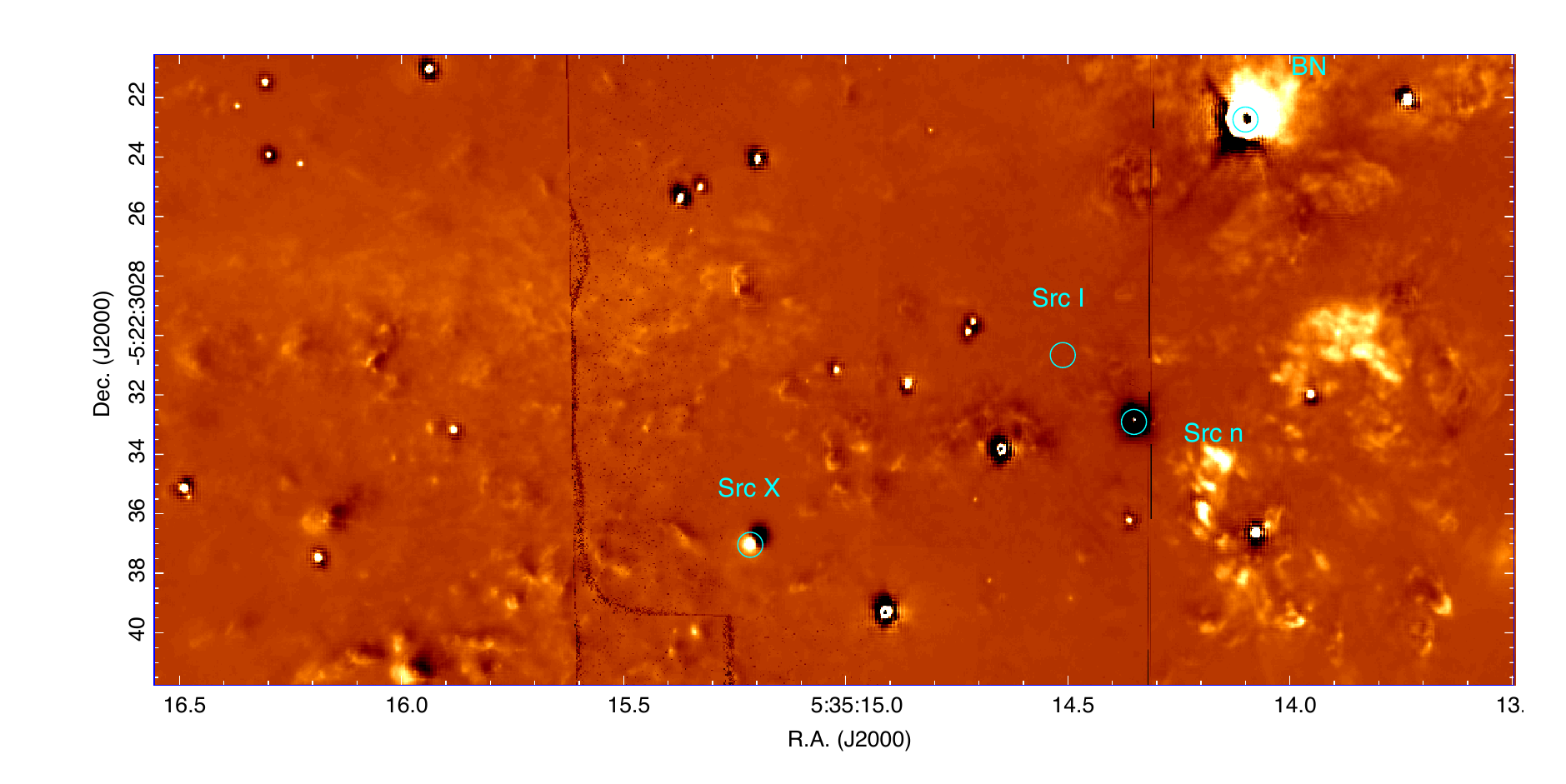}
\caption{A difference image formed from the 2013 epoch  2.12 $\mu$m narrow band \Htwo\ image from \citet{Bally2015} and the 1999 epoch Subaru image from \citet{Kaifu2000}.   
}
\label{fig2}
\end{figure*}

Multi-epoch infrared images obtained at wavelengths ranging from
2 to 12 $\mu$m and listed in Table~1 are used to confirm the 
$\sim$55 \kms\  proper motion of Src x and to place limits on the motion 
of Src n.    The FITS images were  
registered to the  0.06\arcsec\ resolution laser-guide-star adaptive-optics 
2013 epoch image published in \citet{Bally2015} and shown in Figure~\ref{fig1}
using two dozen stars contained within the sub-region of the image shown in 
Figure~\ref{fig2}.   IRAF routines  GEOMAP and GEOTRAN were used to match the 
infrared images listed in Table~1 to the 2013 data.

The $\sim$21 year interval between the \citet{AllenBurton1993} 
images obtained in 1992 and the  \citet{Bally2017} images obtained
in 2013 confirm the high proper motion of Src x.  The \citet{AllenBurton1993}  image
was a mosaic built using a first-generation 58 by 62 pixel InSb array and the image
PSF is about 2 \arcsec\ in diameter.   The proper motion uncertainty is about
20 \kms\ in amplitude and 20\arcdeg\ in position angle.  

The highest quality ground-based images for proper motion determinations
are the 1999 epoch narrow-band \Htwo\ Subaru  image 
from  \citet{Kaifu2000}  which has a 0.5\arcsec\ PSF and the 2013 epoch 
narrow-band \Htwo\ Gemini South  image  from  
\citet{Bally2015} which has a 0.06\arcsec\ PSF.  
Both data sets were built from observations taken over a range of dates;
MJD = 51189 to 51191 for the 1999 epoch Subaru image and
MJD = 56291 to 56351 for the 2013 epoch Gemini South image.  
The time interval between these images  is 5131$\pm$31 days 
($\approx$ 4.43$\times$10$^{8}$ seconds).  Figure~\ref{fig2} shows a difference image 
obtained by subtracting the registered 1999 image from the 2013 
image.

Only a few stars are in common between the \citet{Smith2005} 11.7 $\mu$m image
which has a 0.3\arcsec\ PSF and the \citet{Bally2015} image with a 9 year interval.  However
comparison of the registered images confirm the motion of Src x with a large 
uncertainty in amplitude and position angle of the proper motion vector.

\citet{Dzib2017} list two additional high-velocity sources in their $\sim$5 GHz
radio proper motion analysis, in their table designated Zapata 11, and `IRc23'.
We compared  the 6 GHz radio radio continuum  data from epoch 2012 JVLA observations, 
previously published by \citet{Forbrich2016},  to the above infrared images  
to show that these fast moving features near BN do not contain 
compact sources between  1.6 and 11.7 $\mu$m.  Additionally, we use
the ALMA data obtained under program 2016.1.00165.S to show that
Zapata 11 and IRc23 do not contain compact sources at millimeter wavelengths.   
Thus, as discussed below, these objects are likely to trace ejecta from BN rather
than stars and will therefore not be included in the motion of the center of mass
calculations.

We use 6 GHz JVLA images obtained on 2016 November 27   
\citep{Forbrich2019} together with the 6 GHz VLA data  observed on 
2012 September 30 to October 5  \citep{Forbrich2016} to measure new proper 
motions for the radio sources near BN.    In  2012,  30 hours of on-source data
was acquired over five days in two 1 GHz-wide bands centered at 4.7 and 7.3 GHz.   
The combined  images have an rms noise level of 3 $\mu$Jy~beam$^{-1}$ and a
synthesized beam diameter 0.19\arcsec\ $\times$ 0.30\arcsec\ at P.A.
30\arcdeg .   In 2016,  only  4 hours of on-source data was acquired
in a configuration identical to the one used in 2012. The interval between these 
observations is   
1.31$\times 10^8$~seconds (4.16 years).

\section{Results}

Table~2 lists the motions of the three runaway  stars as well 
as two radio sources  
near BN which were listed in \citet{Dzib2017}
in the Orion reference.  We use a distance of 400 pc 
for the conversion of proper motions (given in milli-arcseconds per year) to 
velocity vectors.     For Src I and BN, the most
reliable proper motion determinations are those of \citet{Rodriguez2017};
for Src x, the best determinations are the HST-based measurements of
\citet{Luhman2017}.   
Table~3 lists what we deem as the best values 
for the motions of the ejected stars.

\subsection{Constraints on Stellar Masses}

We used these measurements and their errors to
compute the motion of the center of mass (CoM) on the plane of the sky
of the three ejected stars as functions of their masses, $M_{BN}$, $M_I$, 
and $M_x$.   If the proper motion vectors of the three stars in the Orion
reference frame are given by
$V_{BN}$, $V_I$, and $V_x$, respectively, the CoM of the system of stars
moves with velocity 
$$
 V_{CoM} = 
 {{(M_{BN} V_{BN} + M_I V_I + M_x V_x ) }
    \over
    {(M_{BN}  + M_I  + M_x  ) }
    }
$$
The masses of BN, Src x, and Src I are  allowed to span a slightly 
wider range than indicated above.   The
proper motions are allowed to vary in both amplitude and position angle
over the range of uncertainties.

It is assumed  that OMC1 is stationary in the reference frame defined by 
stars in Orion Nebula Cluster.  Furthermore, it is assumed that before
the dynamical interaction ejected these runaway stars from OMC1, their
center of mass was moving through OMC1 with a speed
less than the escape speed.  We use the 850 \um\  dust continuum map
of Orion A from \citet{Lane2016} to estimate the mass of OMC1 as a function
of radius from its center using \citet{Lane2016} equation 1.   
The dust temperature as a function of distance from
OMC1 is estimated by averaging the values from 
\citet{Dicker2009}, \citet{DeBuizer2012}, and \citet{Chuss2019}.   
The gas-mass within 0.05 pc of the ejection center is 
$M_{OMC1} \approx$ 100 \Msol\ for  a mean dust temperature $T_d$=60 K. 
The mass inside a circle with a radius of 0.5 pc
increases  to $\sim$1,300 \Msol\ (for $T_d$=30 K).   
At these radii and assuming that the total enclosed mass at a given radius
is dominated by the gas mass, the equivalent circular orbit speeds would be 
$V_{circ}  \sim$2.9 and 3.3 \kms\ at radii of 0.05 and 0.5 pc, respectively.  The
mass distribution is roughly that of an isothermal sphere with a relatively
constant equivalent circular orbit speed.   If the mean
velocity of each star prior to ejection was equal to the equivalent circular orbit 
speed, the center of mass motion is likely to be lower than the circular orbit
speed by $\sqrt 3$.  Thus, the motion of the center of mass is likely to 
be less than $<V_{circ}> / \sqrt 3 \approx$1.9 \kms .   

\begin{figure*}
\gridline{\fig{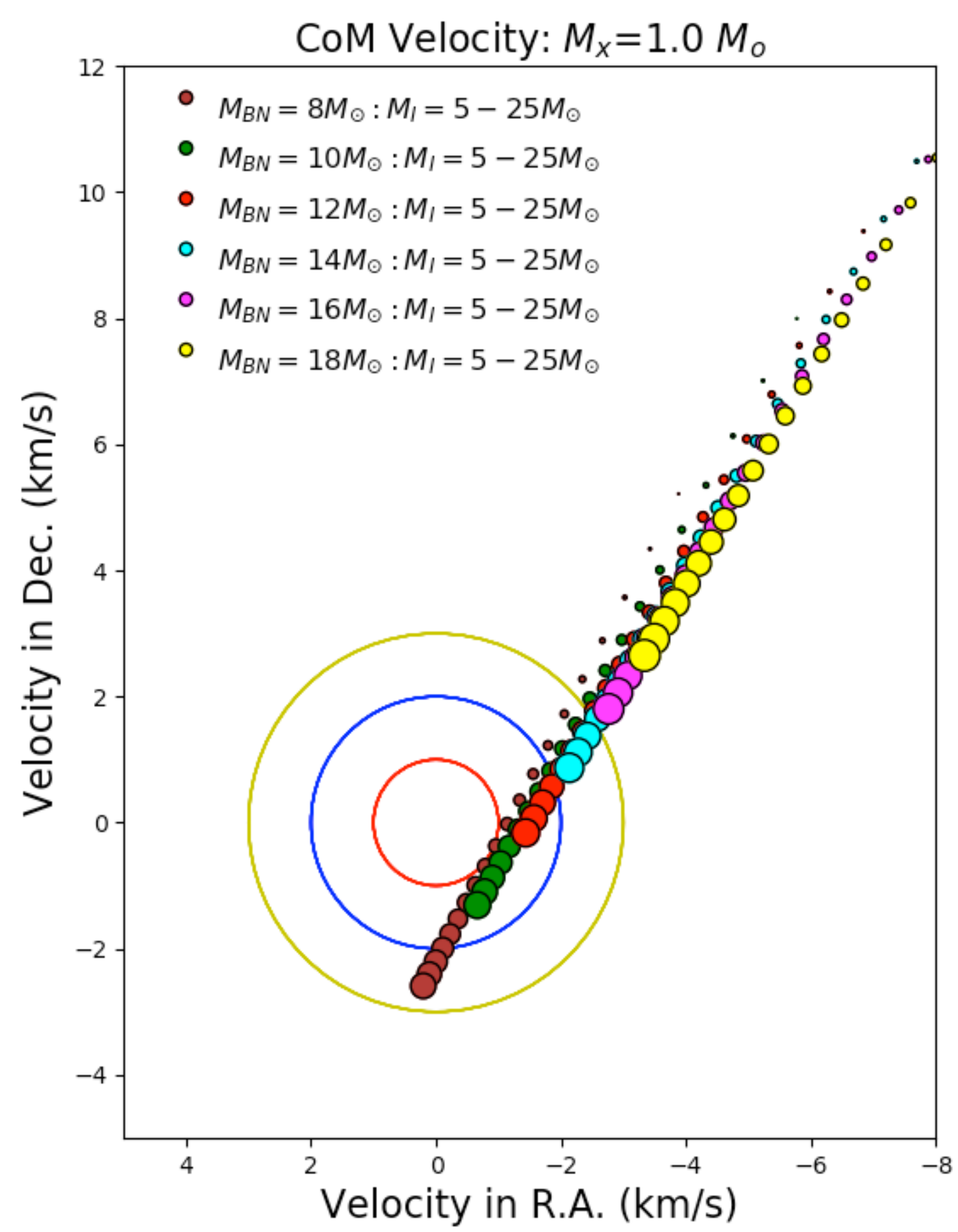}{0.43\textwidth}{(a)}
              \fig{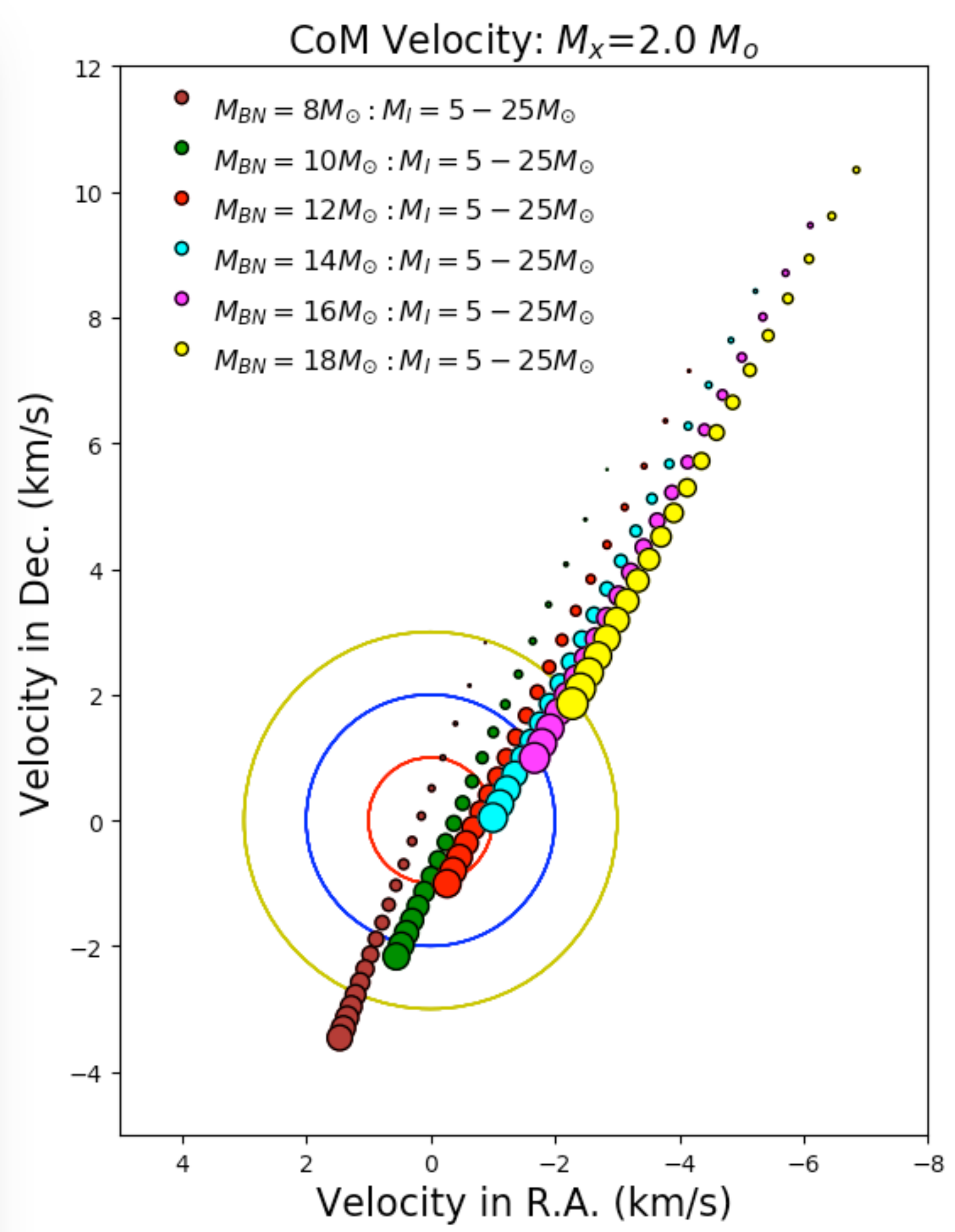}{0.43\textwidth}{(b)}
              }
\gridline{\fig{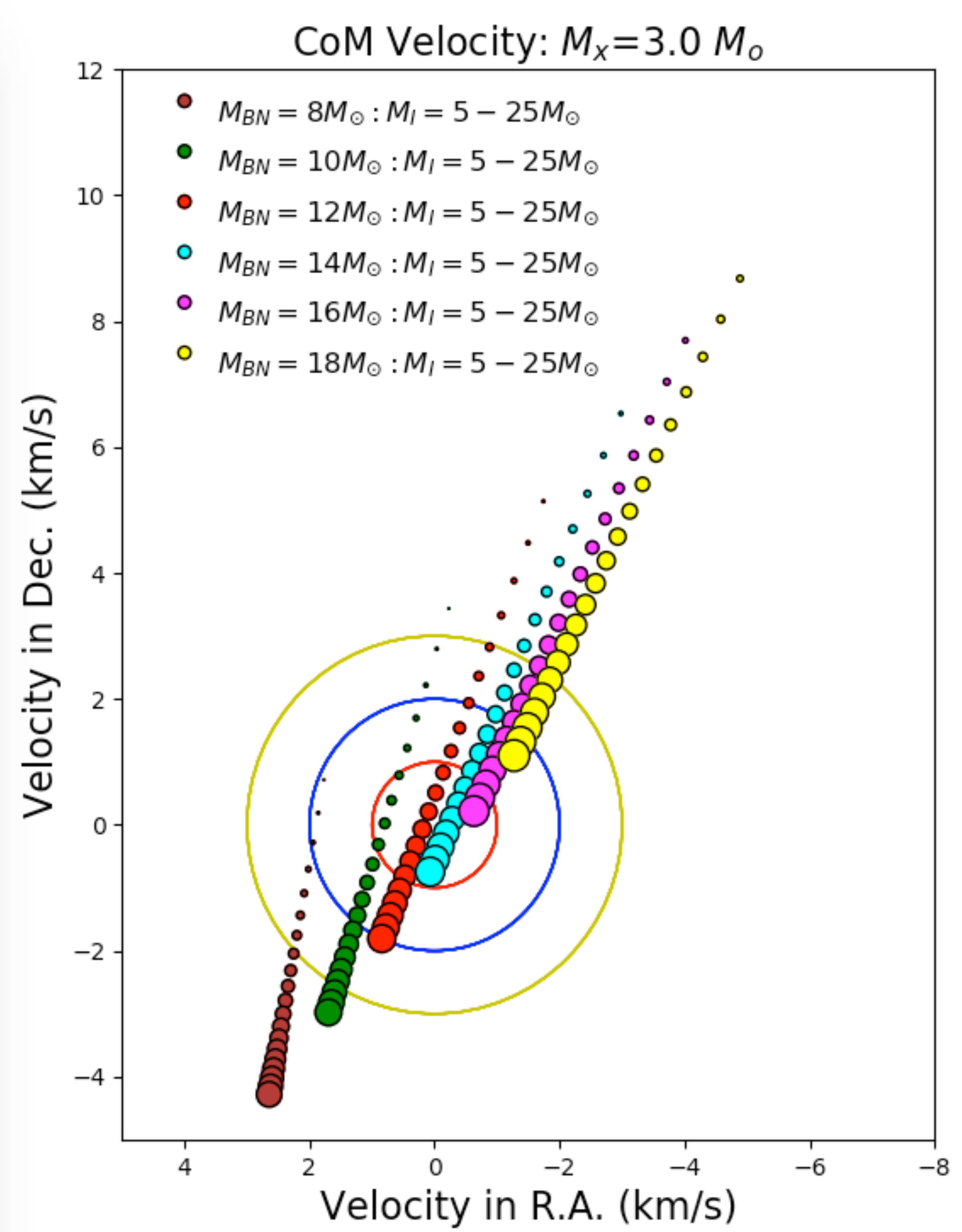}{0.43\textwidth}{(c)}
              \fig{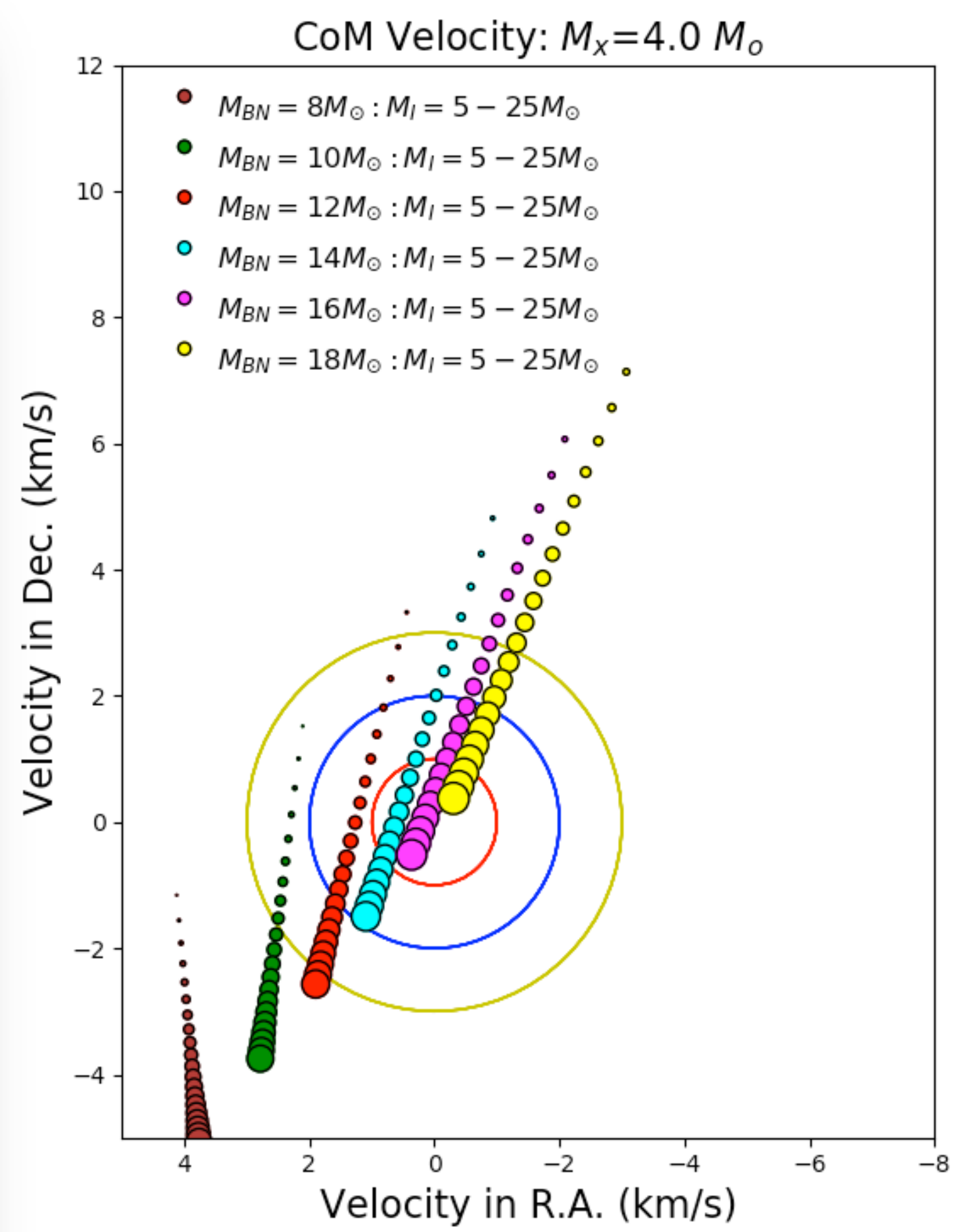}{0.43\textwidth}{(c)}
}
\caption{The velocity of the center of mass (CoM) of BN, Src I, and Src x  as a function of the masses of Src~x, BN, and Src~I.    The position angles and proper motions assumed in this calculation are the best-fit values from \citet{Rodriguez2017,Luhman2017} and assume that IRc23 and Zapata 11 are very low-mass ejecta from BN.  The red, blue, and yellow circles correspond to escape speeds of 1, 2, and 3 \kms\  from OMC1 with respect to the center-of-mass of the Orion Nebula Cluster.  Each panel shows a specific choice for the mass of Src~x.  For each chosen mass of Src x, each panel shows  range of possible masses for BN shown in a unique color.   For each Src~x and BN mass, the mass of Src~I varies from 5 \Msol\ (top right) to 25 \Msol\ (lower left).  The circle diameters scale with the total mass, M$_I$ plus M$_{BN}$. 
\label{fig3}}
\end{figure*}

Figure~\ref{fig3} show the calculated motion of the CoM in the Orion 
reference frame as functions of the stellar masses and proper motions.  
Each panel corresponds
to a specific choice for the mass of Src x, starting with $M_x$ = 1 \Msol\ in the
upper-left.   The mass is increased by 1 \Msol\ in each panel and ends 
with $M_x$ = 4 \Msol\ in the lower-right.  For each choice of $M_x$, the mass
of BN is allowed to vary from $M_{BN}$ = 8 to 18 \Msol\ to span the outer bounds 
of likely values;  each choice of $M_{BN}$ is given a unique color shown in the
legend.  For each choice of $M_x$, and $M_{BN}$, the mass of Src I is allowed
to vary from $M_I$ = 5 to 25 \Msol\ in increments of 1 \Msol\ to bracket the 
full range of plausible masses.  ($M_I$ = 5 \Msol\ corresponds to the top of 
each string of points while $M_I$ = 25 \Msol\ corresponds to the bottom). 

The total mass of the three ejected stars is likely to be larger than $\sim$20
\Msol.   Thus, because of equipartition of  kinetic energy, their center of mass 
is likely to have a motion  with respect to the OMC1 frame that is considerably 
lower than the velocity dispersion of the majority of low mass stars in
the  Orion Nebula Cluster (ONC).   The blue, red, and yellow circles 
in  Figure~\ref{fig3} indicate the speed of the CoM with respect to the 
ONC reference frame  in steps of 1, 2, and 3 \kms.  
The center of these circles is assumed to 
be at rest in the Orion Nebula reference frame determined from the 
mean motion of 79 radio stars which defines a reference frame with 
$\mu _\alpha cos (\delta)$ = 1.07$\pm$0.09 mas~yr$^{-1}$ and
$\mu _\delta $ = $-$0.84$\pm$0.16 mas~yr$^{-1}$ \citep{Dzib2017}.
At a distance D = 400 pc, 1 mas~yr$^{-1}$ corresponds to a motions
of 1.896 \kms\ implying that the velocity dispersion of these radio stars,
1.175 mas~yr$^{-1}$ (taking the geometric mean of the dispersion
in R.A. and Dec. from Dzib et al. 2017), is $\sim$ 2.2 \kms\, slightly
larger than the blue circle in Figure \ref{fig3}.

A low mass Src x shifts the CoM to the northwest.  As the mass of   
Src~x increases, the CoM shifts to the southeast for all likely Src~I and
BN masses. 
Figure \ref{fig3} shows that in order for the center-of-mass of the
Src~I-BN-Src~x system to move with a speed less than $\sim$2 \kms\
in the ONC reference frame, 
a low mass ($\sim$1 \Msol ) for Src~x requires a high mass 
($>$ 11 \Msol ) for Src~I, regardless of the mass of BN.   
On the other hand, a  high mass for Src~x requires a relative high mass 
BN ($>$10 \Msol ).

For the most likely mass of Src~x, 2 to 3 \Msol ,  a low mass BN allows 
a low mass for Src~I.  For a more massive BN  the motion of the CoM 
tends to be away  from the Trapezium towards northwest.  Thus, a higher Src~I 
mass is required to keep the CoM motion at rest with respect to OMC1. 
For a $M_x$=3 \Msol , $M_{BN}$=12 \Msol , $M_I >$ 14 \Msol\  if
the CoM is moving away from the ONC with less than 1 \kms . 
If the CoM of these ejected stars is at rest or falling towards the ONC,
a higher mass for Src I is more likely.

Variation of the proper-motion speeds and position-angles over the
range allowed for by the uncertainties listed in Table~1 were used
to study their impact on the estimated motion of the center of mass 
of the three stars.  Varying the velocity and position angle of each 
star over the full range of allowed values while holding the corresponding
parameters at their central values results in an error circle for the
CoM motion.  The radius of this error circle is about 0.4 \kms , 
nearly independent over the range of stellar masses.   Allowing all 
three speeds and position angles to vary over their full range of acceptable
values simultaneously results in an error circle with a radius of about
0.6 \kms , again relatively constant over the full range of stellar masses
under consideration.    As shown by Figure~\ref{fig3}, the largest uncertainty
is caused by the relatively  poorly constrained mass of Src x.

\subsection{Constraints on the Radial Velocity of Src x}

Although the radial velocity of Src x has not been measured, it
can  be estimated from momentum conservation combined with the 
radial velocities and masses  of BN and Src I from 
$$
V_r(x) = - ~ {{M_I~V_r(I) + M_{BN}~V_r(BN)} \over {M_x}}
$$  
Assuming that $\rm V_r(BN)=+14.2 \pm 0.5$~\kms\  
      and                  $\rm V_r(I)    =-4\pm0.5$~\kms\  in the OMC1 
reference frame,  that  the mass of Src x is 
                              $\rm M_x = 3 \pm 0.5 $~\Msol , 
                              $\rm M_{BN} = 12 \pm 2$~\Msol , and
                              $\rm M_I = 15 \pm 2$~\Msol\ 
 implies  $\rm V_r(x) = -37 \pm 10$~\kms .  Thus, the
 LSR radial velocity of Src~x is predicted to be
 $\rm V_{LSR} = -28 \pm 10$~\kms .

The study presented here differs from that of \citet{FariasTan2018} 
who also used the motion of the center of mass of the ejected
stars in the Orion frame to constrain the stellar masses in several respects.  
Our study explores a wider range of masses for Src x.    We use the
measured radial velocities of OMC1, BN, and Src I to constrain
the likely radial velocity of Src x.   Our analysis is broadly consistent
with the conclusions of \citet{FariasTan2018}.  We present a graphical illustration 
of how this constrain implies a mass for Src I around 15 \Msol , consistent
with the direct mass measurement presented by \citet{Ginsburg2018}.

Prior to the discovery of the large proper motion of Src x by \citet{Luhman2017}, 
source n was though to be a runaway star ejected ~550 years ago.   
As noted by \citet{Bally2017}, it was not possible to reconcile the
stellar proper motions with the requirement that the 
stellar center of mass is stationary in the ONC reference frame.  Indeed, the
motions seemed to suggest that CoM and OMC1 might be moving with a velocity 
of a few kilometers per second towards the west with respect to the 
ONC.    The discovery that source n is not moving, but Src x is,  resolves
this dilemma.   Reasonable choices for the masses of the stars allows
for no proper motion in the ONC frame.  

It is possible that the mutual gravitational attraction between the ONC
and OMC1 actually results in a net proper motion of OMC1 and the CoM
towards the southeast where the center of mass of the ONC cluster
lies (very close to the Trapezium stars, located $\sim$90\arcsec\ from OMC1).
As shown in Figure~\ref{fig3}, a motion of the CoM and OMC1 towards the Trapezium 
would favor a higher mass for both Src I and BN.    
This point was also discussed by \citet{FariasTan2018}.

\subsection{A Slow Radio Jet from BN?}

\begin{figure*}
\fig{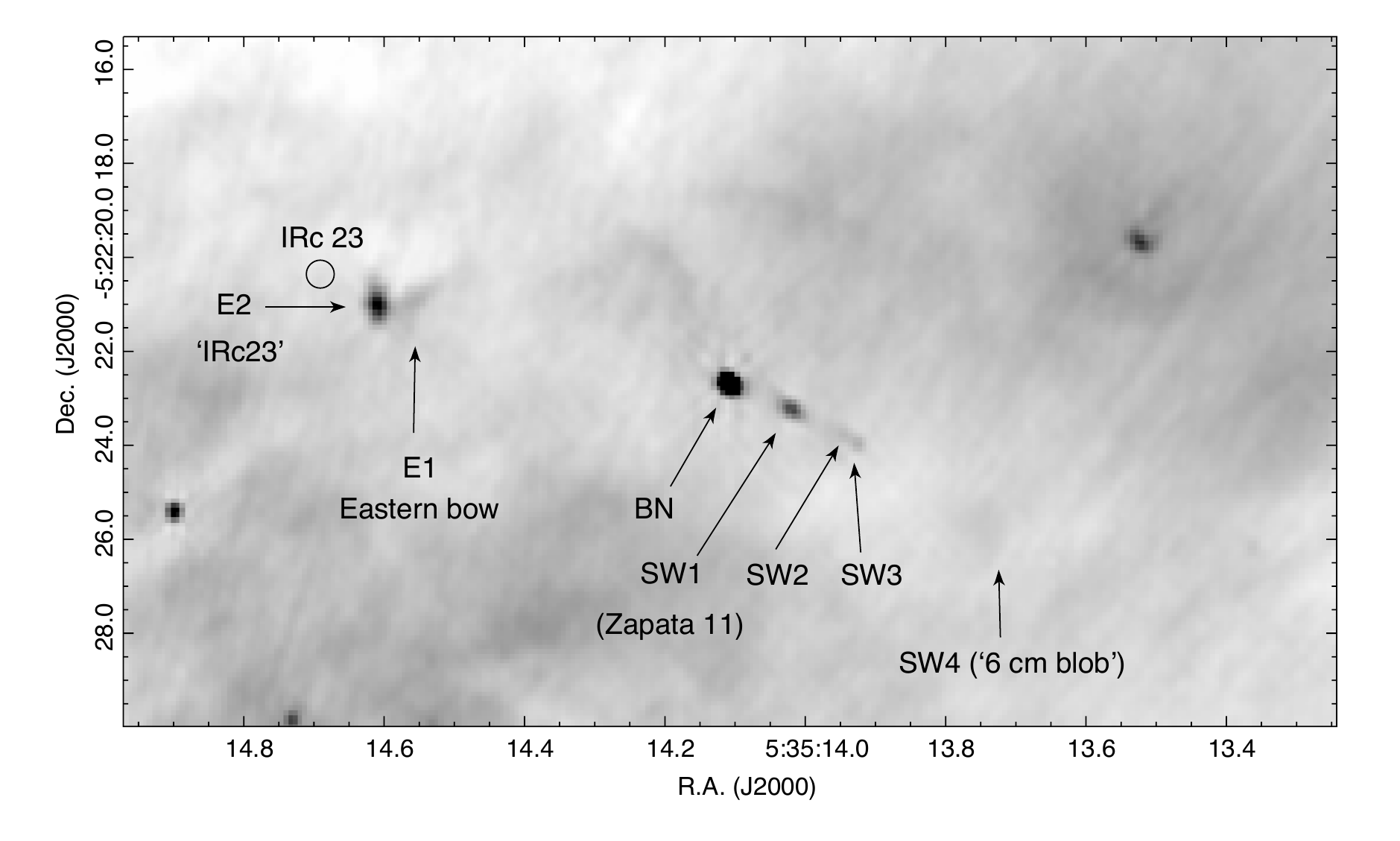}{1.05\textwidth}{}
\caption{A close-up  showing a small portion of the 6 GHz  radio continuum image from \citet{Forbrich2016} with the various objects
listed in Table~4 indicated. 
}
\label{fig4}
\end{figure*}

\begin{figure*}
\fig{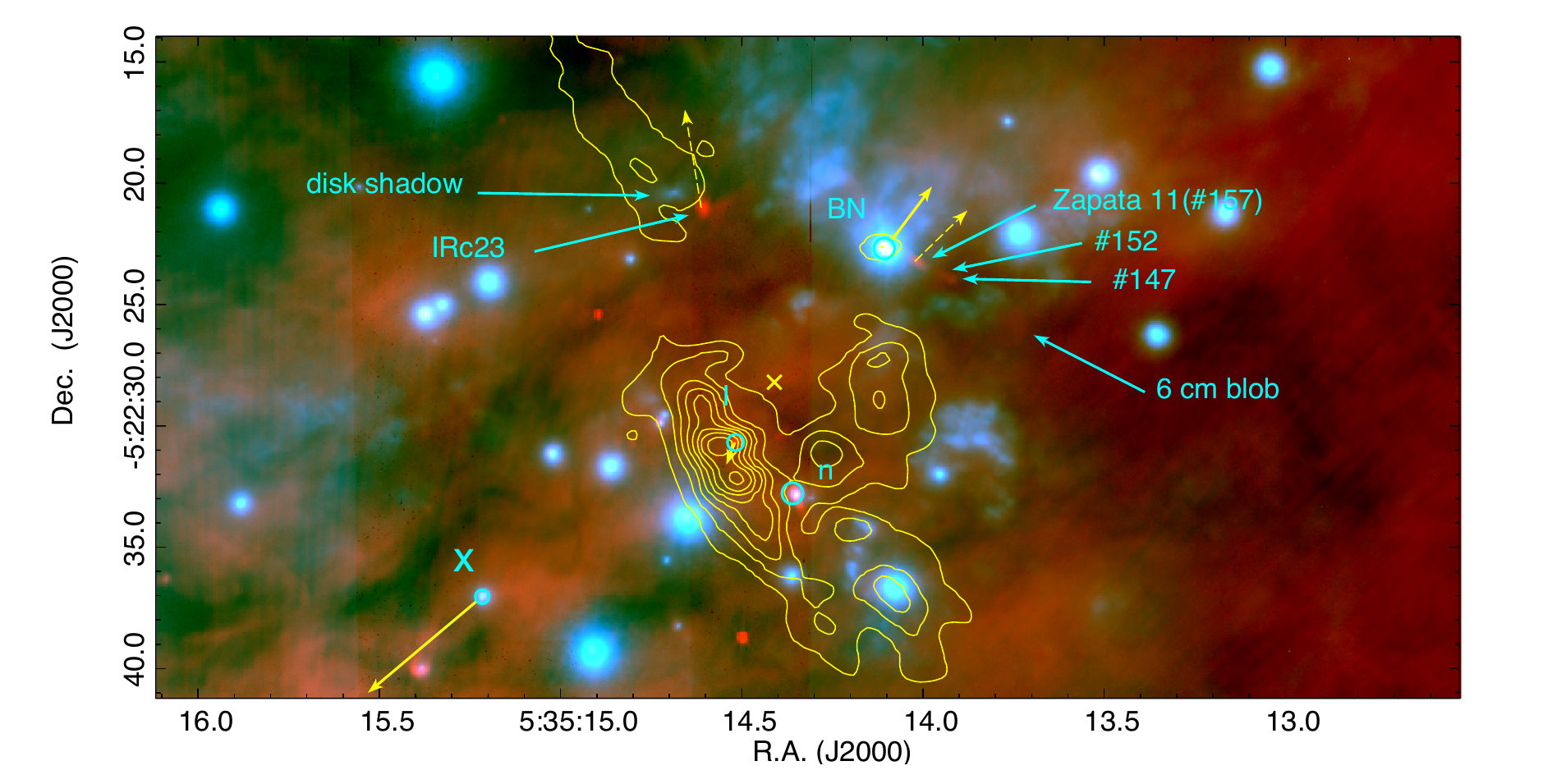}{1.05\textwidth}{}
\caption{A color composite image showing a portion of the 6 GHz radio continuum from \citet{Forbrich2016} in red, superimposed on the 1.64 $\mu$m [Fe~II] image 
from \citet{Bally2015} in  cyan.  
Yellow contours show the 1.3 mm dust continuum emission from OMC1 at 1\arcsec\ resolution \citep{Bally2017} with contour levels at 
50, 156, 261, 367, 472, 578, 683, 689, 789, and 894  mJy/beam (the beam has a FWHM of about 1\arcsec ).  
The solid yellow arrows mark the proper motions of the ejected stars with lengths corresponding
to the motion over the next 200 years.   Dashed arrows show the proper motions of E2 (`IRc23') and SW1 (`Zapata11') taken from  \citet{Dzib2017}.  The `X' marks the 
center of the explosion as determined by \citet{Bally2011,Bally2015,Bally2017}.  The radio emission feature labeled `IRc23' lies
at the western edge of the cometary clump of dust pointing roughly to the explosion center.   A compact bipolar reflection nebula with
an axis at PA$\sim$100\arcdeg\ shows the location of a `disk shadow', presumably illuminated by an obscured central star located in the head
of the cometary dust feature.   The numbered 6 GHz sources (\#152 and \#147) southwest of BN were tabulated by \citet{Forbrich2016}.  The  `6 cm
blob' is an extended enhancement of radio continuum emission along the axis defined by Zapata 11, \#152, and \#147.
}
\label{fig5}
\end{figure*}

Figure \ref{fig4} shows a closeup of the region around BN in the 2012 epoch 
JVLA 6 GHz   data from \citet{Forbrich2016}.   BN is surrounded by a chain of 
diffuse  radio sources consisting of SW4 (6 cm blob), SW3, SW2, and SW1 
(Zapata 11 in Dzib et al. 2017)  located southwest of  BN and E1 and E2 
(`IRc23' in Dzib et al. 2017)  
located east of BN.     These objects trace a gently curving chain bending back 
toward the explosion center.    The designations of these
features in \citet{Forbrich2016} are indicated in Figure \ref{fig5}.
We consider three possible interpretations for these radio features near BN.
  
First, as  suggested by \citet{Dzib2017}, they may contain low-luminosity protostars.
If stars were embedded in the compact radio sources,  the mass moving towards 
the north and northwest is larger than  just BN.    
In this case, the mass of Src~I needs to be increased to keep the CoM less than
the escape speed from OMC1.   However, there is  no evidence  
for  X-ray, infrared,  or sub-mm counterparts to any of these radio features.  

Second, given the apparent motion of at least two of the features, Zapata 11
and `IRc23' \citep{Dzib2017} away from the explosion
center, and the point from which BN, Src I, and Src x were ejected, 
the radio features may trace clumps ejected during the dynamic interaction 
and explosion $\sim$550 years ago.  Their apparent association with BN is 
either a coincidence, or indicates that they are clumps which became 
unbound  from BN during its acceleration.  

Third, they may trace a low-velocity outflow from BN.   
SW1 (Zapata 11; Zapata et al. 2004) and E2 `IRc23'; Dzib et al. 2017)  
exhibit proper motions away from the explosion center as 
well as away from BN (Figure \ref{fig5}).   These objects are spatially extended
in the 6 GHz radio continuum.   E2  is the brightest knot in the chain around
BN.   New radio proper motions are presented below and listed in Table~2.

The radio source dubbed  `IRc23' by  \citet{Dzib2017} but  designated 
E2 here, is offset by more than 1\arcsec\ from the  mid-IR source IRc23 
first  identified on 12.5 $\mu$m images  \citep{Shuping2004}.  
At 11.7 $\mu$m  \citep{Smith2005},  IRc23 is a condensation along the eastern 
edge of the mid-IR nebula centered on BN that abuts the 
western edge of a  cometary dust cloud seen at millimeter wavelengths 
(Figure~\ref{fig5}).    

A  `disk shadow' (Figure \ref{fig5}) and bipolar reflection nebula is 
centered on IRc23 at the southwest-facing  head of the cometary cloud.    
This nebula indicates the presence of a star embedded in a nearly edge-on 
disk with an axis at  PA $\sim$ 105$\pm$5\arcdeg\ (disk shadow is elongated towards 
PA = $\sim$15\arcdeg ).   A 16-th magnitude star (Vega scale) is visible at the 
center of the  `disk-shadow'  in the $K_s$  adaptive optics images 
\citep{Bally2015}.   No compact 1.3 to 3 mm source brighter than 0.1 mJy/beam
is seen in the 0.03\arcsec\  resolution ALMA continuum  images
associated with data  presented by \citet{Ginsburg2018,Ginsburg2019}.  
 
E2 (designated \#68 in Forbich et al. 2016 and `IRc23' in Dzib et al. 2017), 
is an arcsecond-scale 6 GHz 
knot  elongated north to south about  1.3\arcsec\  southwest of  
IRc23 (Figure \ref{fig5}).   
A  dim, 2\arcsec\  long, crescent-shaped of  radio feature, E1, is 
located about 0.5\arcsec\  west from E2.  E1 resembles a bow 
shock moving due-east from BN.    E2 and
E1 may  trace an outflow running into  the 
western rim of the cometary dust cloud.  

SW1  (designated \#157 in Forbrich et al. 2016 and Zapata 11 in Dzib et al. 2017)
is the brightest  knot in a jet-like,  chain of radio continuum 
emission extending southwest from BN containing SW2 and SW3  
(Forbrich et al. 2016 objects \#152,  \#147)
which terminates in a very dim blob, SW4 (`6 cm blob' in Figure \ref{fig5}).
 
None of these radio features contain compact 
1.3 mm sources in the ALMA continuum image.   
The 1.3 mm flux limit is $\sim$0.1 mJy in a 0.03\arcsec\ beam.
In the 11.7 $\mu$m image  of \citet{Smith2005} there are no 
compact counterparts  brighter than 0.3 Jy at the locations
of E1 and E2,  SW1, SW2, SW3, or SW4.
However, there is a local intensity maximum near SW1.    
In the 12.5 $\mu$m  image \citep{Shuping2004} this feature is labeled `SW Arc'   
and appears to be part of the bright mid-IR nebulosity around BN.
In the K$_s$ adaptive optics image from \citet{Bally2015}, 
there are no compact ($<$0.1\arcsec ) sources brighter than magnitude 
18 at the locations of the 6 GHz knots.  
However, bright,  diffuse emission due to reflection nebulosity and
H$_2$ emission limits sensitivity.
\citet{Grosso2005} used the Chandra X-ray observatory to observe 
OMC1 between 0.5 and 8 keV for an on-source exposure of 838 ks (9.7 days).
There are no compact X-ray sources at the locations of these radio features.  
The limits on the X-ray luminosity are L$_x < 10^{27}$ erg~s$^{-1}$.  

Assuming that the string of knots near BN trace an outflow, the  mean position
angle of this flow on the plane of the sky is PA$\sim$69\arcdeg .
Multiple knots in a chain indicate that the outflow is collimated into
a velocity-variable beam with a width of hundreds of AU.

\subsection{New Proper Motions and Variations of the 6 GHz Features Near BN}

We use the 2012 and 2016 VLA data to measure new proper motions and to identify
variations in the radio emission. 
\footnote{A full discussion of proper motions derived from a comparison of these two datasets will appear in Vargas-Gonzalez et al., in prep.} 
Table~2 lists these measurements; Table~4 
gives these proper motions relative to the motion of BN.    SW1, SW3, and E2
are moving away from BN with speeds of 28$\pm$9 to 45$\pm$20 \kms .   SW3 brightened from
a peak flux of $\sim$80~$\mu$Jy in 2012 to $\sim$300~$\mu$Jy in 2016 and SW2 
increased from $\sim$60~$\mu$Jy to $\sim$300~$\mu$Jy.   Additionally, the structure 
of knots  SW2 and E1  changed significantly over the 4 year interval.    
The photo-centers of these features seems to be moving back  towards BN!     
A possible interpretation is that SW2 on the 2012 image merged with SW3 by  2016, resulting 
in the more than factor three increase in the flux of SW3.  Additionally, a new knot  
has appeared in 2016 slightly east of where  SW2 was located in 2012.   
The diffuse feature SW4 was not recovered in the shallower 2016 epoch images.

The photo-center of the northwest  edge of E1,  the `eastern bow' 
located $\sim$1\arcsec\ west of E2,  shifted towards the northwest.     
Fitting the photo-centers gives an apparent proper motion of $\sim$300~\kms .
However,  it is more likely that
this `motion'  is caused by intensity variations; fading of one part of the 
`eastern bow' combined with brightening of another part.      Given the errors, 
the motion of BN  is consistent with previously measured proper  motions 
using longer time intervals.

\subsection{Electron Densities of the 6 GHz Features Near BN}

Table~5 lists the background-subtracted,  peak and area-integrated flux-densities,  
the radii of the regions in which the flux-densities were measured, 
the  projected distance from BN, and the estimated mean electron density
in each knot in Figure \ref{fig4}  measured on the  2012 epoch
6 GHz image.  The spectral index of the knots between  4.7 and 7.3 GHz
is flat.  Thus, we assume that the radio continuum emission is produced by
optically thin bremsstrahlung.

The optical depth of free-free emission is given by 
$\tau = 3.28 \times 10^{-7} T_4^{-1.35} \nu _1^{-2.1} EM $
\citep{CondonRansom2016} where $T_4$ is the plasma temperature in
units of $10^4$ K, $\nu _1$ is the frequency in units of 1 GHz, 
$EM = \int _0 ^L n^2_e(x) dx \approx  n^2_e ~ L~~(cm^{-6} pc)$ 
is the emission measure in the usual units, $n_e$ is the electron density, and
$L$ is the line-of-sight path-length through the emission region (in units of pc).  
Thus, $n_e \approx (EM / L)^{1/2}$ (cm$^{-3}$).
When $\tau <<$ 1, the flux-density is
given by $S_{\nu} = 2 k T \nu ^2 \Omega _B (sr) \tau / c^2$.  
Thus,
$$
S_{\nu}(Jy)  = 100.77 ~T_4^{-0.35}~ \nu _1^{-0.1}~ \Omega _B(sr) ~EM
$$
For the JVLA 6 GHz observations the synthesized beam diameter is approximately
0.36\arcsec , implying that $\Omega _B = 2.39 \times 10^{-12}$ sr.    
With $\nu$ = 6.099 GHz,  
$EM(cm^{-6} pc) = 4971 ~S_{\nu}  (\mu Jy)$
where the flux-density is expressed in units of $\mu Jy$.  The mean electron
density is then given by
$$
n_e \approx \Big[ {{4971  ~S_{\nu} (\mu Jy)}  \over {L_{pc}}} \Big]^{1/2} 
~~~~~~~(cm^{-3})
$$
The electron densities given in Table~5 assume that for non-round  features,
the line-of-sight depth is comparable to the minor axis size of the region 
under consideration.   Thus the electron densities range from $10^4$ to about
$10^5$ cm$^{-3}$, about one to two orders of magnitude higher density than
the mean electron density of the Orion Nebula.   If these knots are only partially
ionized, their hydrogen densities will be higher. 

Based on the proper motion measurements, SW1, SW3, and E2 are moving away 
from BN (Table~4).   The lower-bound on 
the masses of the features range from $10^{-7}$ \Msol\ to $10^{-5}$ \Msol .  
Their momenta are at least 
$10^{-5}$~\Msol ~\kms\ to   $3\times 10^{-4}$~\Msol ~\kms ; lower bounds on
their kinetic energies in the BN frame  are $10^{39}$ to $10^{41}$ erg.
For a mean speed of 35 \kms , their  dynamic ages, $\tau _{dyn} = R / V$, range 
from  $\sim$80 years (SW1) to 400 years for E2. 

\section{Discussion}

\subsection{Consequences for the Motion of the Center of Mass}

The lack of compact, stellar sources in the 5 GHz knots  implies that they  
be ignored in the analysis of the motion of the center of mass of ejected stars.  
However, if they did contain significant mass such as a brown dwarf or a star, 
then the mass of Src I  must be increased to keep the motion of 
the center of mass of the ejected objects smaller than the velocity dispersion of OMC1.

\begin{figure*}
\plotone{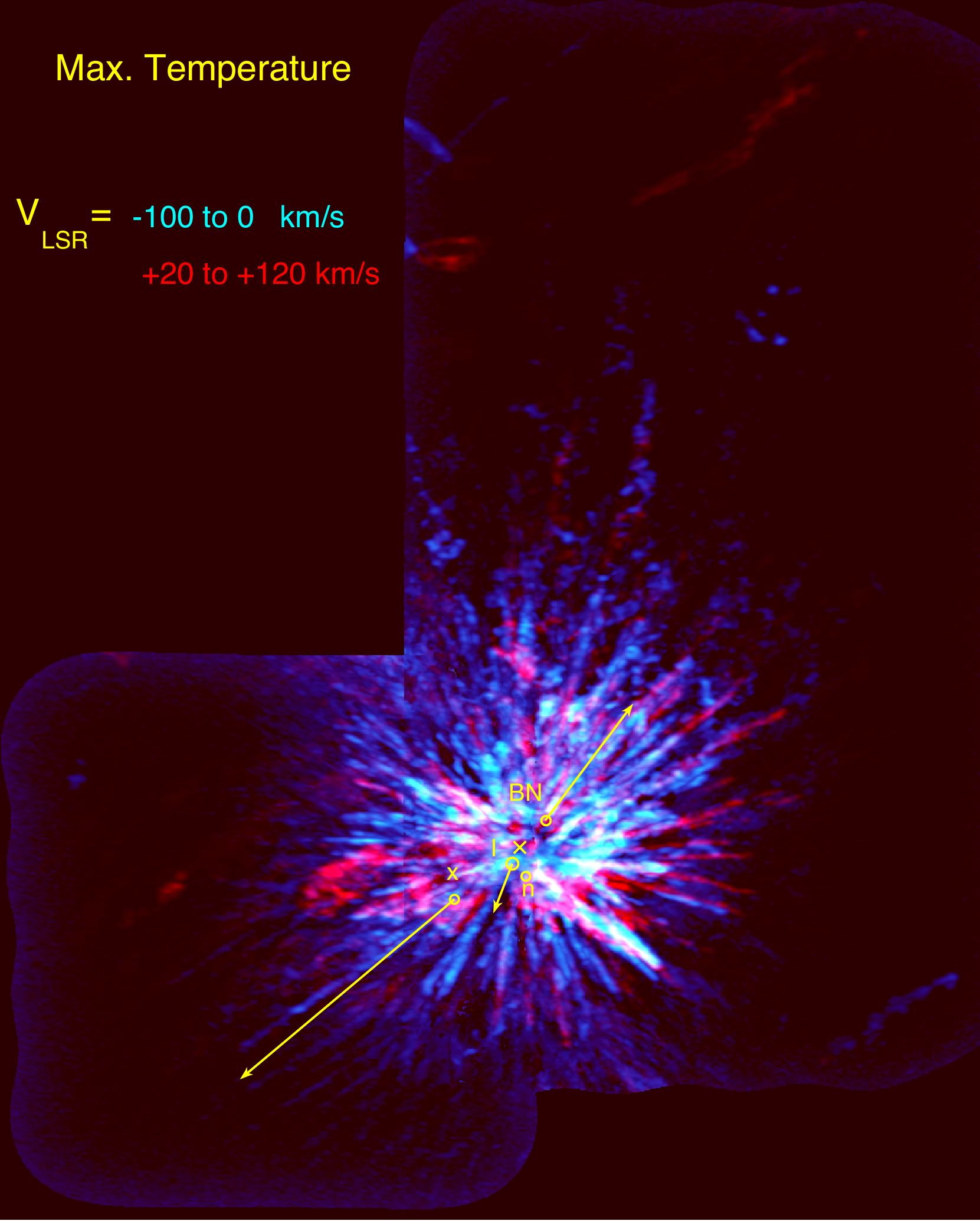}
\caption{A CO J=2$-$1 image showing the Orion OMC1 explosion with 1\arcsec\ angular resolution from  \citet{Bally2017} showing  Src x, Src I, BN, and Src n.   The yellow arrows show the proper motions of the three ejected stars with lengths corresponding to the motions over the next 2,000 years.  The vectors shown are based on the motions measured by \citet{Rodriguez2017} for the BN object and Src I, and by \citet{Luhman2017} for Src x.  
}
\label{fig6}
\end{figure*}

A somewhat greater uncertainty is introduced by the mass and momentum of the
ejected gas in the OMC1 high-velocity outflow and explosion (Figure~\ref{fig6}).  
However, as discussed
in \citet{Bally2017}, 1\arcsec\ resolution observations of the CO in the flow 
with ALMA show that to first order, the explosion in the gas is isotropic.  To second 
order, the flow is elongated along a southeast-northwest axis with a slight deflection 
of the southeastern lobe towards the east.   These deviations from perfect symmetry 
are evident in Figures~\ref{fig1} and \ref{fig6}.    Because the flow is 
roughly elongated in a direction  parallel to the proper motions of BN and Src I, this 
asymmetry is likely to only have a minor impact on motion of the center of 
mass of the ejected stars.

\subsection{Constraints on the BN Disk }

Near-IR 1.6 $\mu$m images of BN in polarized light reveal a 
bipolar reflection  nebula with an axis of symmetry at PA $\approx$ 36\arcdeg\  
\citep{Jiang2005}.   The dark lane in the nebula is close to  
BN's proper motion vector.   If this feature caused by shadowing
by a disk surrounding BN, the disk axis must also be nearly orthogonal 
to the BN proper motion vector.   The mass of such a 
disk is difficult to measure because the mm and sub-mm wavelength
emission is dominated by free-free rather than dust emission \citep{Plambeck2013}.  
The mm-wave emission from BN is compact, with a de-convolved major axis
of 0.06\arcsec\ in \citet{Plambeck2013}, about a factor of three smaller than
the major axis of the Src I disk.  In the ALMA 1.3 mm 0.03\arcsec\ resolution
continuum image, the de-convolved diameter of BN is about 0.053\arcsec\
($\sim$21 AU) with a slight east-west elongation implying a radius of about 
10 AU.  
\citet{OtterGinsburg2019} used a more careful analysis 
found a de-convolved size of 0.045\arcsec\  by 0.039\arcsec\ (16 by 18 AU)
with a major axis PA $\sim$60\arcdeg\ in the ALMA band 6 (1.3 mm) image.  
The elongation is along the axis of symmetry of the BN reflection nebula and 
may trace ionized gas rather than a disk.  However, this emission likely places an 
upper bound on the disk radius.

\citet{Scoville1979} found that the CO overtone bands around 2.3 \um\ 
are in emission from BN.   They interpreted this emission hot CO 
(T $\sim$ 3,000$-$5,000 K)  excited by collisions with \Htwo\ 
in a high density environment (n$(H_2) > 10^{10}$ cm$^{-3}$) within
$\sim$1 AU of BN.  

\citet{Indriolo2018} obtained sensitive  mid-IR spectra around 6~\um\ 
toward BN and  failed to detect either H$_2$O or CO in absorption toward 
the star near its radial velocity of $\sim$ 20 \kms .    The absence of 
absorption towards BN at these velocities implies that the disk seen in
emission in the vibrationally excited lines of CO can't 
be edge-on.    The near-IR bi-polar nebula, combined with the 
CO vibrational-band results implies a significant tilt of the disk-axis
with respect to the line of sight so that most of the extinction and
absorption towards BN must arise from the foreground cloud between BN
and the Orion Nebula ionization front.

\subsection{The Outflow from BN}

BN is moving through the molecular cloud behind the Orion Nebula between the
dense filaments and cores traced by the dust continuum (Fig \ref{fig5}).   Thus,
the compact 6 GHz radio features are likely to be propagating into this 
inter-filament molecular  gas whose density is poorly constrained.   
Assuming that BN is 0.1 to 0.2 pc behind the Orion Nebula's
ionization front and using an extinction $A_V \sim$25 magnitudes towards BN
\citep{Scoville1983} implies a mean
density along the line-of-sight to BN of $n(H_2) \sim$ 3 to 6 $\times 10^4$~\cmq ,
the same order of magnitude as the electron densities of the radio knots.  

What ionizes the knots?  Observations of the free-free and H-recombination 
line emission from BN indicates that ionizing, Lyman continuum radiation
from BN is confined to within 20 AU.  Thus, ionization of these radio
features is likely caused by strong shocks.  To ionize hydrogen, the shocks must
be fast enough to either produce Lyman continuum emission which ionizes the 
pre-shock medium in a radiative pre-cursor, or be sufficiently hot to collisionally
ionize hydrogen in the post-shock layer.    Dissociation of   \Htwo\ into
\Hi\ followed by ionization of  at least 50\% of \Hi\  requires a  post-shock
temperature of  $T_{ps} \sim \Phi / 10 k >$15,000 K where $\Phi$ = 13.6 eV is
the ionization potential of  \Hi\   \citep{Draine2011}.  
 
Thus, significant collisional ionization of hydrogen requires a shock speed 
$V_{ps} > (16 \bar n k T_{ps} /  3 \mu m_H)^{1/2} \sim$ 30 \kms .   
Here, $\mu \approx$2.4 is the mean molecular weight
in the pre-shock molecular cloud (assuming Z=0.70, Y=0.28, and Z=0.02), 
$\bar n$ is the number of particles in the  post-shock medium created 
by dissociation and ionization per incident  \Htwo\ molecule (4 particles  when 
 \Htwo\ is dissociated and fully ionized).  For 50\% ionization of H, $\bar n \sim 3$.
 Given  that the BN disk is likely to be inclined, the 
measured proper motions imply that the  actual shock speeds are likely to 
be $\sim$30 to 50 \kms .

The recombination time for  a plasma with electron density $n_e$ is
$t_{rec} = 1 /  n_e \alpha _B$.   For $n_e = 10^4$ to $10^5$~cm$^{-3}$
$t_{rec}$= 1.2 to 12 years.  Thus, it is not surprising that large photometric 
variation  are observed on a time-scale of $\sim$ 4 years, or that the sizes
of the shocks in the BN outflow have sub-arcsecond dimensions along their 
minor axes. 
 
High spectral resolution observations of the near-IR,  hydrogen recombination 
lines provide evidence for a  dense, ionized  wind emanating from BN with 
$\dot M \sim 4 \times 10^{-7}$~\Msol ~$yr^{-1}$ \citep{Scoville1983}.
\citet{Beuther2010} presented higher signal-to-noise data confirming 
the broad H-recombination lines with a FWHM of $\sim$175 \kms\ 
along with a brighter compact line-core with FWHM of $\sim$39.3 \kms .
The wind remains ionized to a radius of $\sim$20 AU from 
BN where it has a density  $n_e \sim 1.3 \times 10^7$~\cmq .   Best fits to
the Brackett~$\alpha$ and $\gamma$ line profiles indicate a  wind velocity profile 
that declines as $V_w = V_o (r / r_0)^{-2/3}$ where $r_0 $= 0.025 AU 
\citep{Scoville1983}.     For an initial wind-velocity of $1$~to~$2 \times 10^3$  
\kms\ at the base of the wind, assumed to be located at the r = 0.025 AU stellar radius
of a ZAMS B0 star,  
this velocity profile implies that the wind decelerates to 11 to 23 \kms\ at r = 20 AU . 
However, slight deviations from the above power-law wind velocity profile can result
in a higher wind-speed.
Proper motions and shock-ionization of the knots indicate that
the flow speed must be at least 30 to 50 \kms\ at distances of a few hundred AU
from BN. Our observations indicate that on larger scales, this wind is  collimated into
a velocity-variable jet emerging nearly orthogonal to BN's compact circumstellar
disk and the star's proper motion.

\subsection{Dynamical Ejection and Disk Properties}

Src I is surrounded by a massive ($\sim$0.1 \Msol ),  
nearly edge-on, $\sim$50 AU radius 
disk  whose spin axis is nearly orthogonal  to the Src I proper motion vector 
\citep{Ginsburg2018,Ginsburg2019,Hirota2017}.   The Src I spin axis is at 
PA $\approx$ 53\arcdeg .    
The BN object is also surrounded by $<$20 AU disk.   Its spin axis
is also nearly at right angles to the proper motion as indicated by its
near-infrared reflection bebula.   Both Src~I and BN have 
younger than 500 year old outflows ejected roughly at right angles to 
their proper motions.

Src x has near- to mid-IR excess, indicating a warm and likely compact 
circumstellar component.   However,  Src x is not detected at mm 
or sub-mm wavelengths.  Thus,  any disk associated with this source 
must have a much smaller mass than the disks surrounding Src I and BN.

Close periastron passages during the final N-body interaction leading to 
stellar ejection can severely perturb or disrupt pre-existing circumstellar disks 
and envelopes.       
The abrupt acceleration during a close N-body encounter that imparts 
a velocity $V$ to a star will tend to stripoff and eject  circumstellar 
matter beyond  the gravitational radius,  $R_G \approx GM / V^2$ 
where $M$ is the mass of the accelerated star.  On the other hand, 
material located inside  $R_G$ will tend to be dragged along because it is
sufficiently bound by the star's gravity (Kepler speed is greater than the
star's kick speed).

For a Src~I with $M_I$ = 15 \Msol\   and $V_I$ = 10.2 \kms , 
$R_G \approx$ 128 AU.     The measured radius of the Src I disk ranges from 
$R_I$ = 37 to 51 AU \citep{Ginsburg2018},  well within the current gravitational 
radius of Src I.    For BN with $M_{BN}$ = 12 \Msol\ and $V_{BN}$ = 30 \kms ,
$R_G \approx$ 12 AU, similar to the observed radius of the sub-mm emission
from BN.  For Src~x with $M_x$ = 3 \Msol\ and $V_x$ = 55 \kms ,
$R_G \approx$ 0.9 AU, consistent with the lack of detection of a disk but the presence
of bright near- to mid-IR emission.   

Multiple encounters prior to 
stellar ejection can remove additional matter from a disk or envelope.
A rough estimate of the  fractional area of a disk impacted by a passing star with 
mass $M_2$ moving with speed $V_2$ in the potential of  a more massive star
with mass $M_1$ can be obtained from the gravitational radius of $M_2$, 
$R_{2}(r)  \approx GM_2/ V^2_2(r)$   where  $r$ is the periastron separation when
the intruder star plunges through the disk.  Matter closer to the intruder than  $R_2(r)$ 
is likely to be ejected from the disk around $M_1$  or become captured 
by the intruder.   If the intruder also has a disk, the cross-section for interaction can
be larger.   

Because the intruder star is most likely to have been born from the same clump
as the primary, its speed during disk impact, $V_2(r)$,  can be approximated 
by the escape speed at $r$ from the gravity well of the primary with mass  $M_1$.   
(The intruder star is  accelerated by the gravitational potential of the primary.)  
Thus, $r_G \approx  {{r ~M_2 }  /  {2 ~M_1 }}$.   

If the disk outer-radius is $r_d$,  the fraction of the disk area impacted by an intruder
impacting the disk at right angles is $f \approx (r_G / r_d)^2$, or
$$
f \approx {1 \over 4 }~\Big({{r} \over {r_d}}\Big)^2  \Big({{M_2} \over {M_1}} \Big)^2
$$
which is a lower-bound. As the orbit plane of the intruder approached the disk 
plane,  the impacted area increases because area defined by $r_G$ becomes
an ellipse with a semi-major axis $a_G \approx r_G / sin (\theta)$ where $\theta$
is the angle between the two planes.  The most destructive encounter would be 
one in which the orbit plane of the secondary lies within the plane of the primary's disk. 
Multiple encounters prior to the final stellar ejection can progressively remove ever
larger portions of the disk, leaving retained material in a highly perturbed state.
Each time an intruder star passes through a disk, its likely to eject some mass. 

The massive disk around Src I is likely to be the surviving remnant  of a primordial 
disk that pre-dated the dynamic ejection and explosion $\sim$500 years ago.
If more than one star was surrounded by a disk,  material
could have been exchanged or caught by the object which became  Src I. 
Similarly, parts of the Src I disk might have been trapped by the other ejected stars.  
The retained parts of disks would have been highly perturbed and shocked.  
The event likely heated the entire  disk \citep{Ginsburg2019}.    However,  
the non-circular and non-coplanar motions  would be damped by shocks 
within a few orbits.   At a radius of 50 AU, the orbit time around a 15 \Msol\
Src I is $\sim$100 years, much shorter than the $\sim$550 years since 
the ejection of Src~I.    Thus, its not surprising that  today, the Src I disk looks 
relaxed.   

\begin{figure*}
\gridline{\fig{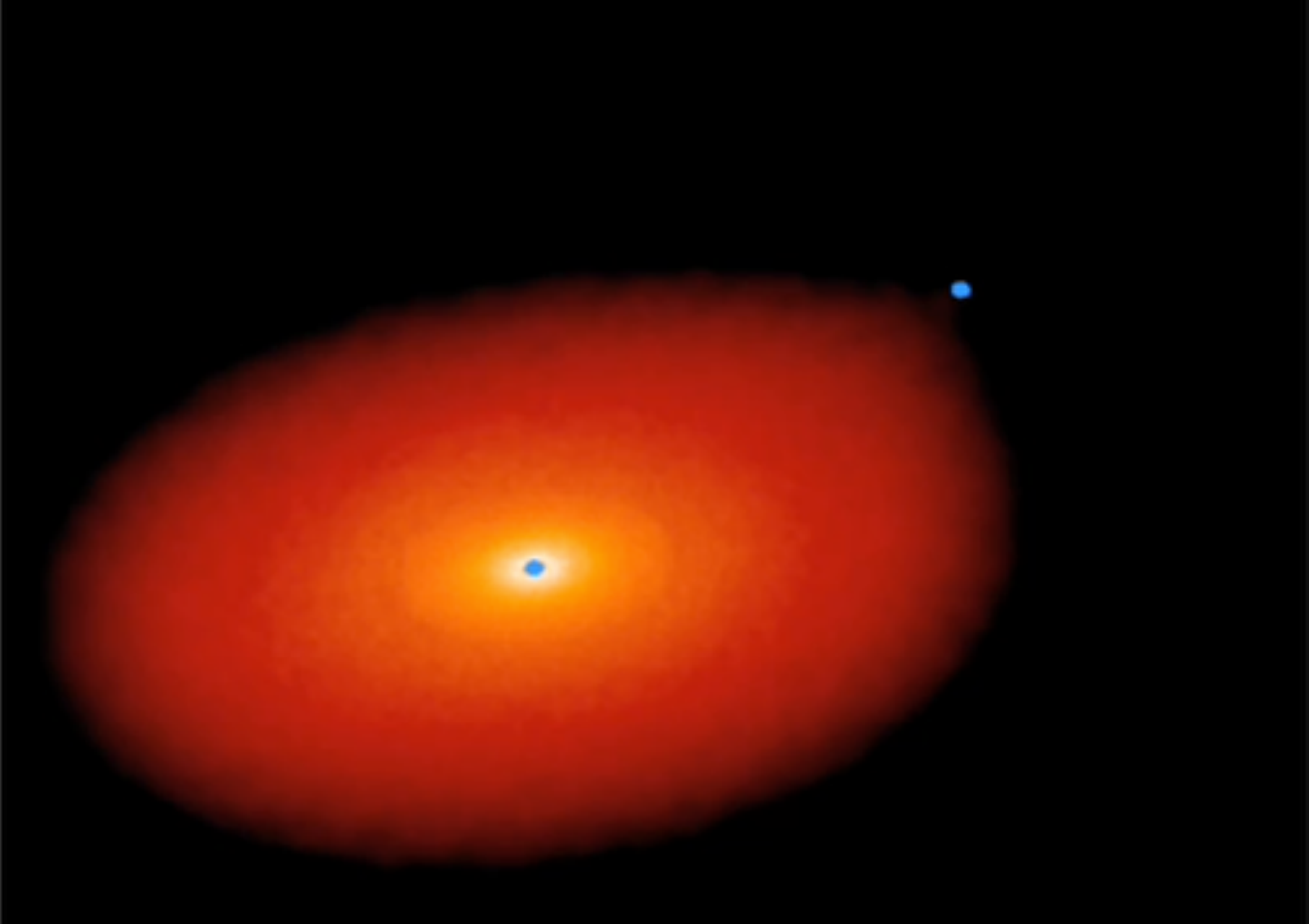}{0.33\textwidth}{(a)}
              \fig{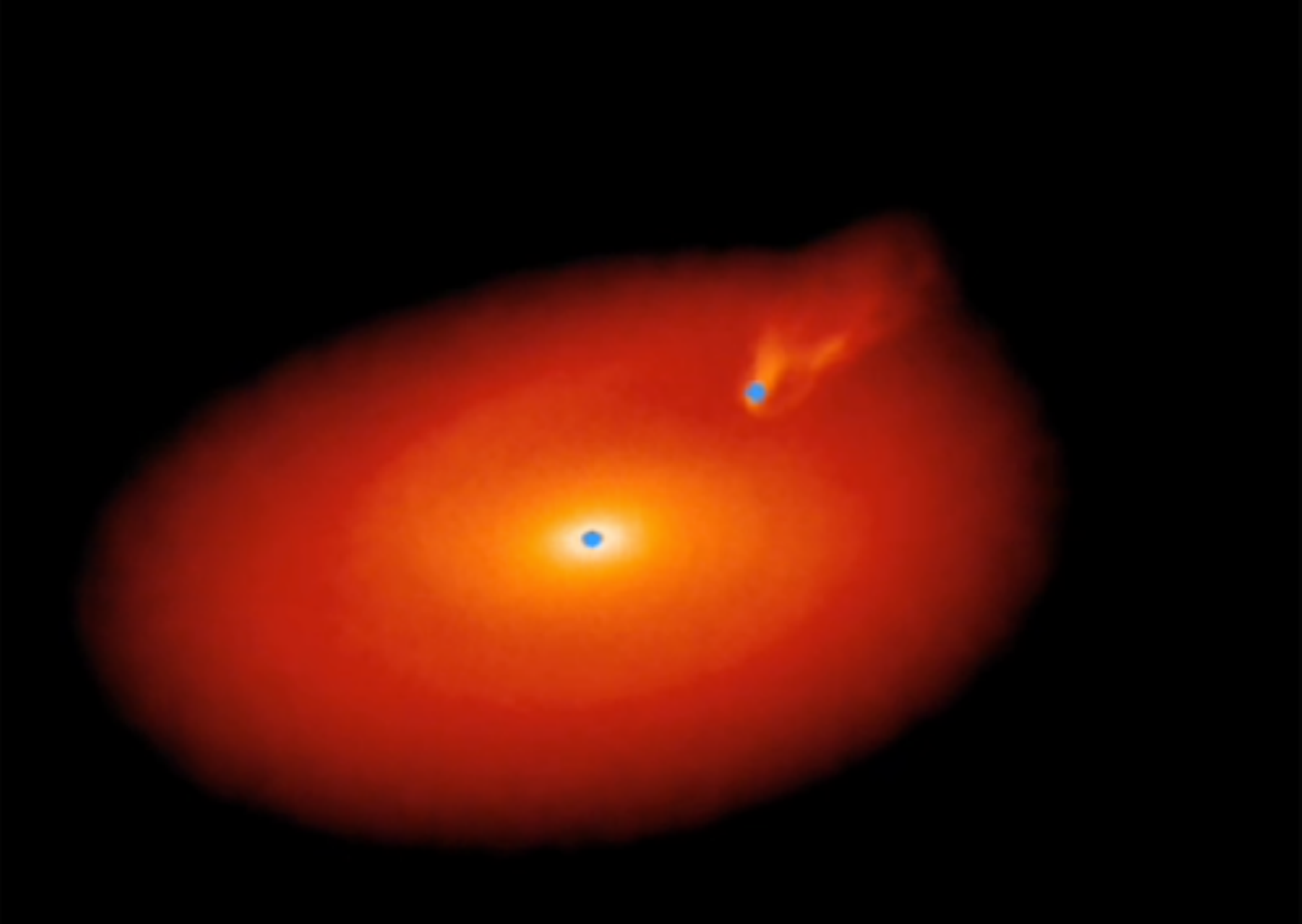}{0.33\textwidth}{(b)}
              \fig{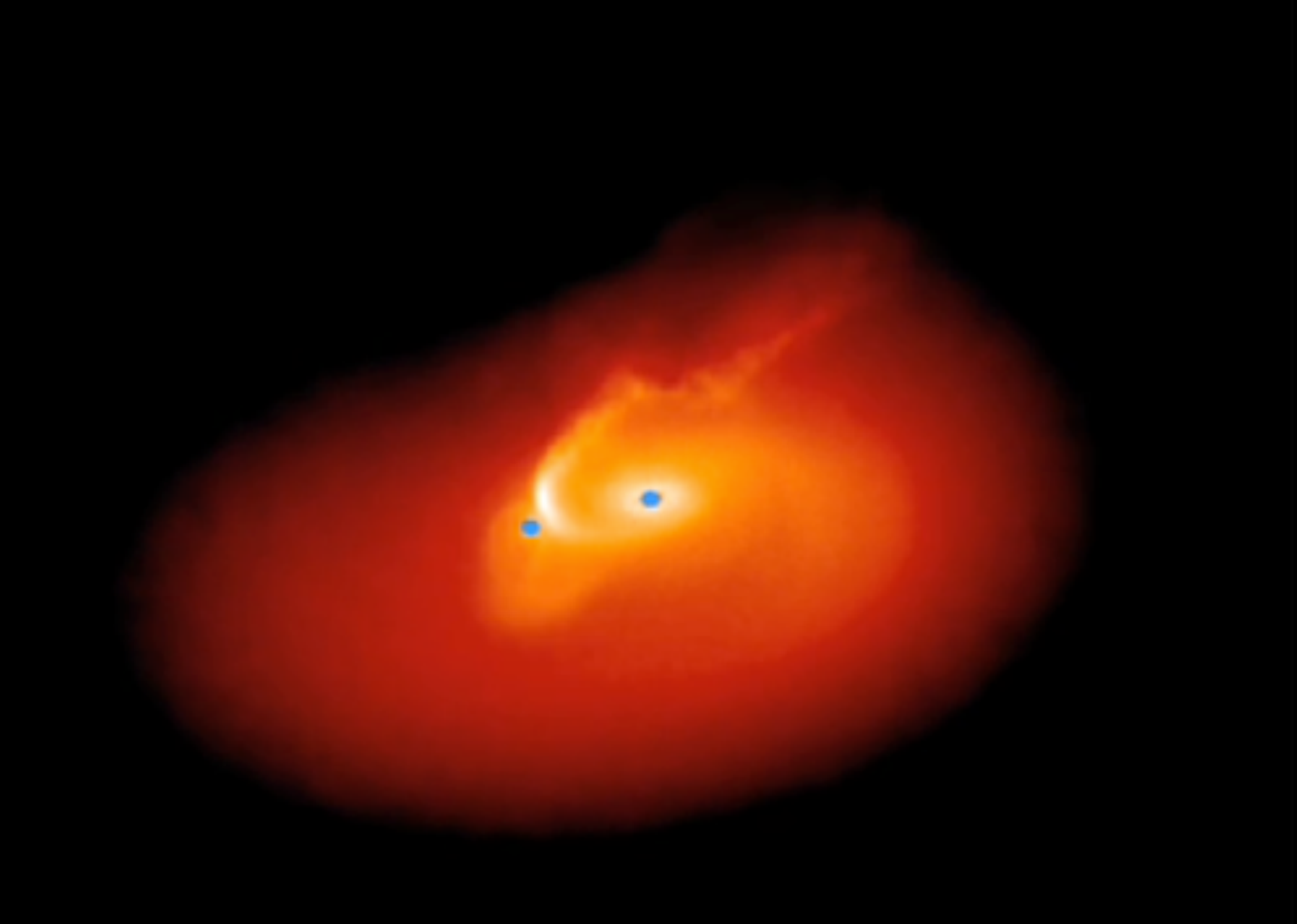}{0.33\textwidth}{(c)}
               }
\gridline{\fig{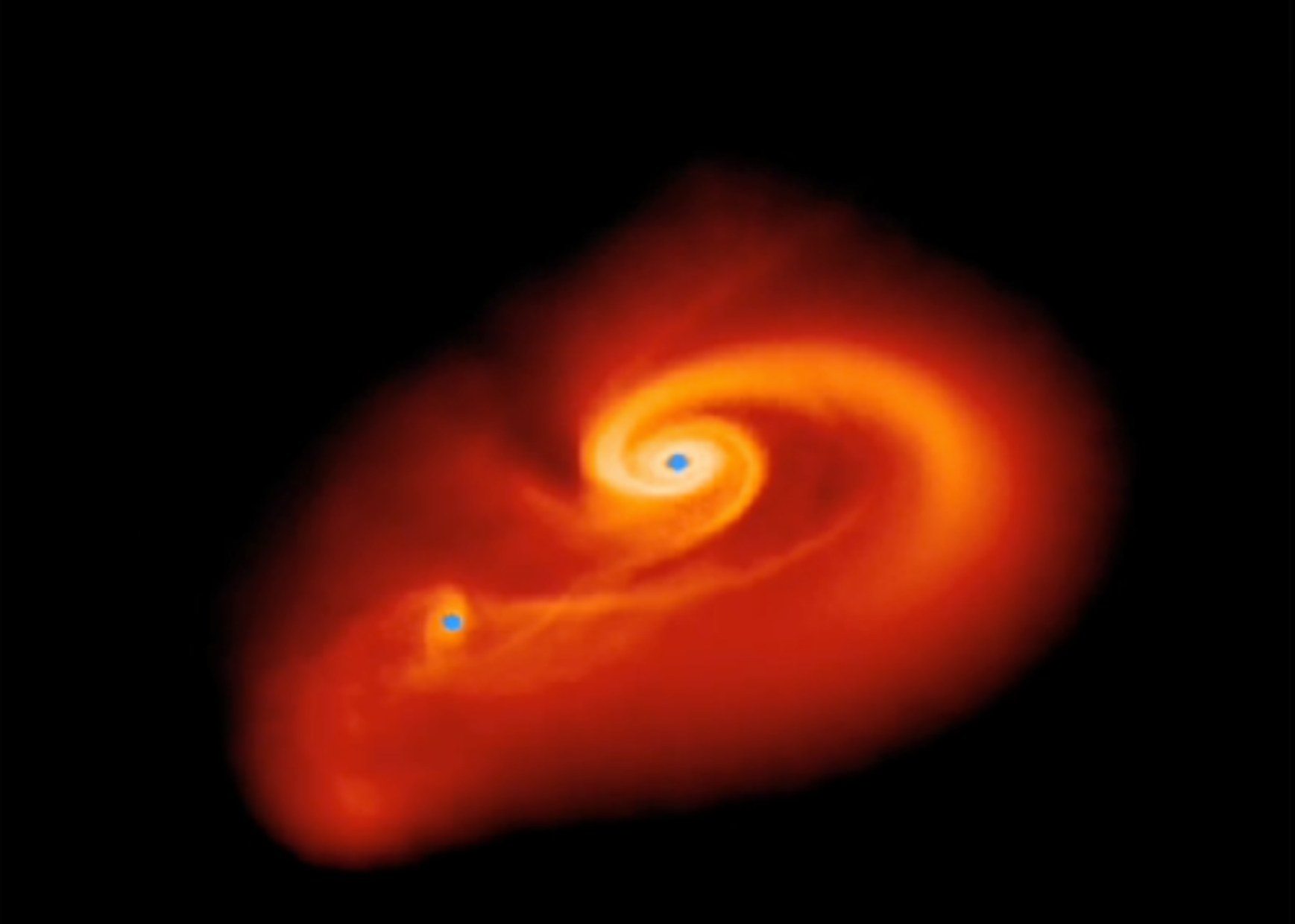}{0.33\textwidth}{(d)}
              \fig{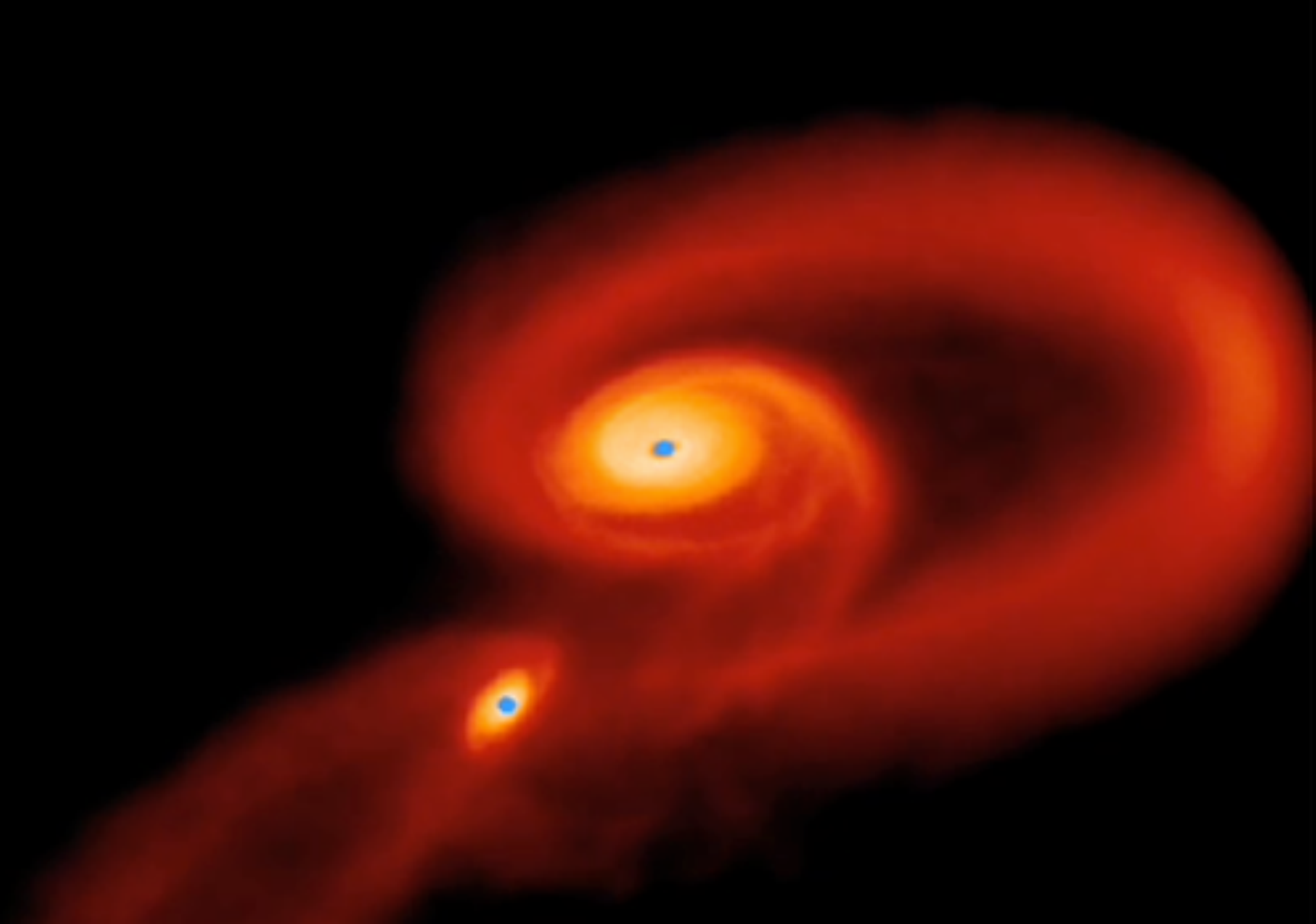}{0.33\textwidth}{(e)}
              \fig{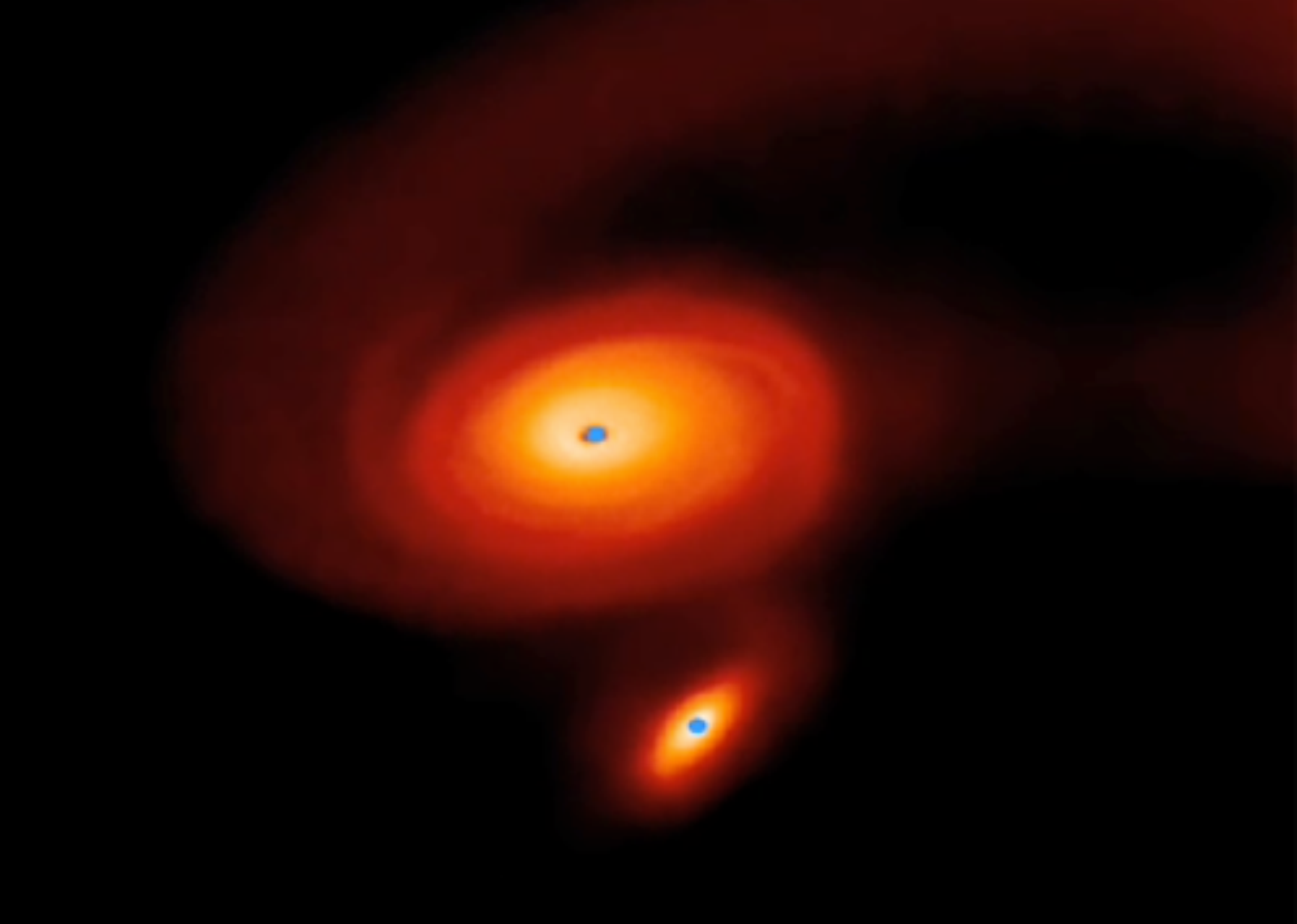}{0.33\textwidth}{(f)}
              }
\gridline{\fig{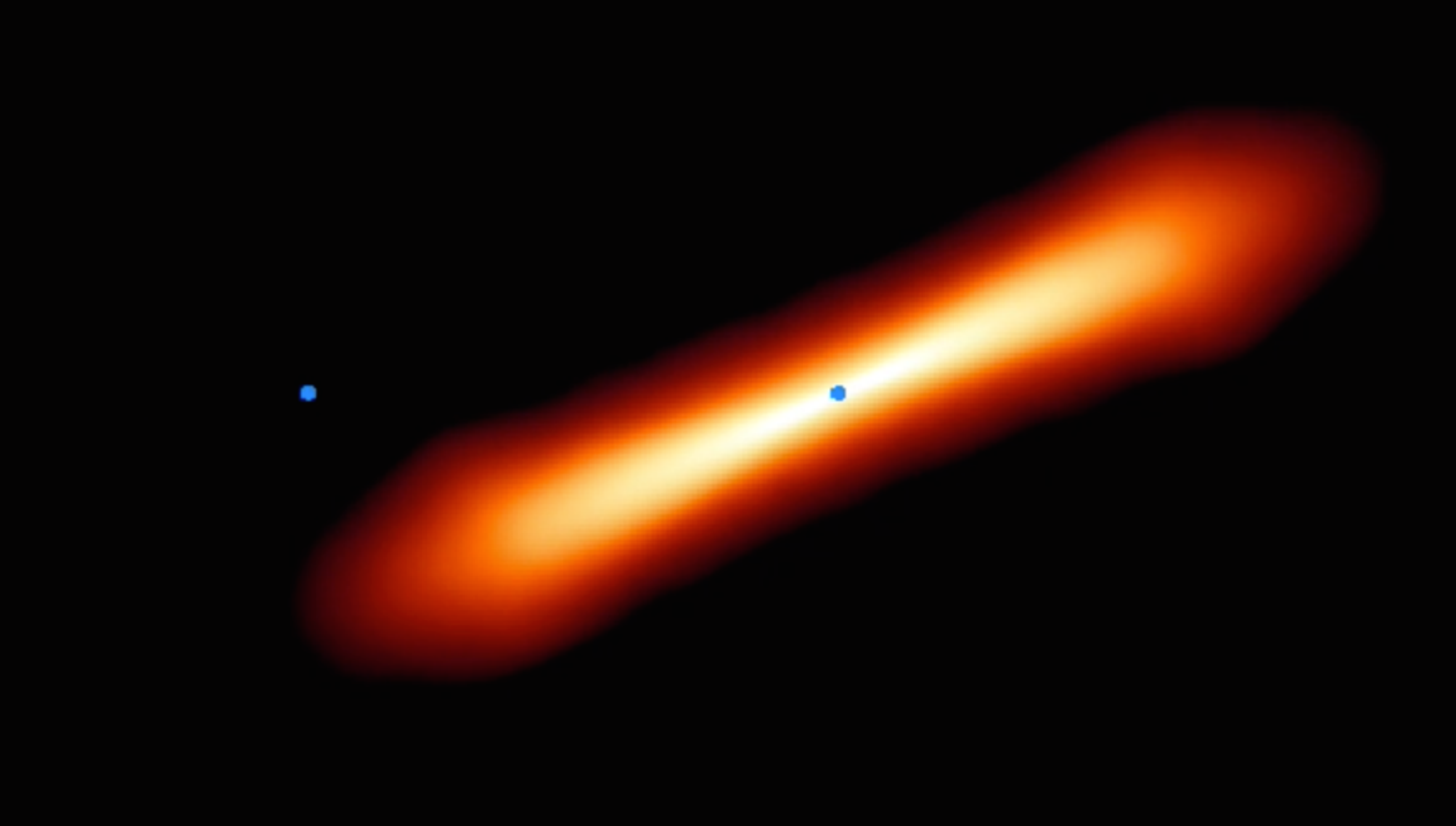}{0.33\textwidth}{(g)}
              \fig{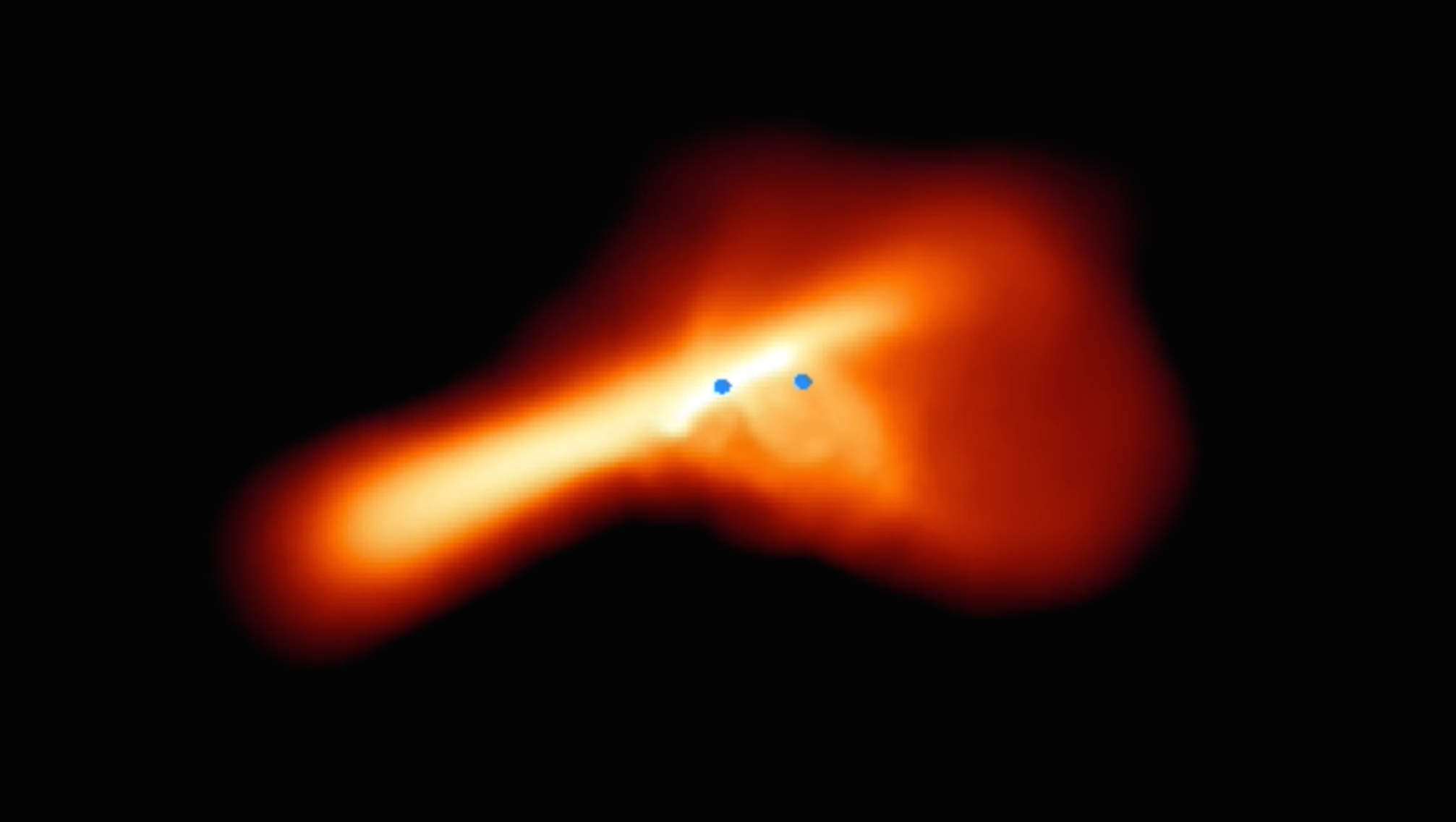}{0.33\textwidth}{(h)}
              \fig{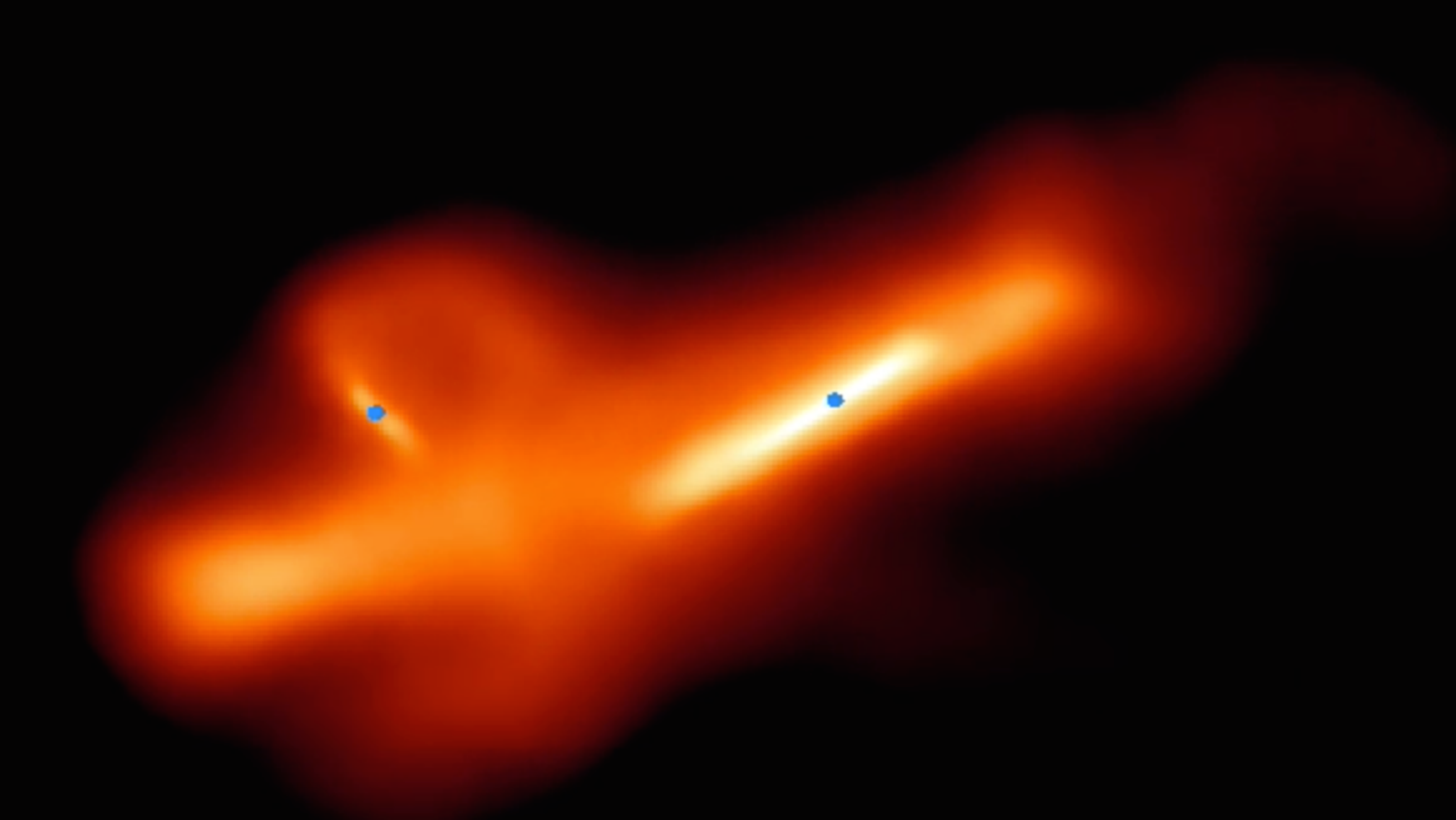}{0.33\textwidth}{(i)}
              }
\caption{{\bf (a) through (f)}:  Snapshots of an sph simulation showing the evolution of the first encounter of a 5 \Msol\ intruder star with no disk in an initially hyperbolic orbit inclined by 30\arcdeg\ with respect to the disk-plane around a 15\Msol\ star.  Orbital energy dissipation traps the intruder into an inclined, elliptical orbit around the primary.  The total mass of the disk is 1.5 \Msol .   The gravitational radius of the 5\Msol\ intruder star is $r_G \approx 0.17 r$. where $r$ is the distance from the 15\Msol\ star.  Frames taken from  simulations used in \citet{Cunningham2009}, \citet{MoeckelBally2007}, and \citet{MoeckelBally2006}.
{\bf (g) through (i)}:   
Frames (a), (c), and (e)  above seen from a vantage point located to the right in in frames (a) through (f) and lying in the initial disk plane of  the massive star. 
\label{fig7}}
\end{figure*}

Figure \ref{fig7} shows several frames from a simulation
of a penetrating encounter 
\citep{Cunningham2009,MoeckelBally2007,MoeckelBally2006}.  
In this 3D SPH simulation, 
a 15 \Msol\ star is surrounded by a massive, 1.5 \Msol\ disk with a surface density 
declining as $r^{-1}$ and  temperature declining as $r^{-1/2}$.   A naked, 5 \Msol\ 
star on a hyperbolic orbit with an orbit plane inclined  30\arcdeg\ with respect 
to the disk plane of the primary encounters the disk (Figure \ref{fig7}a).   During early phases of
the  encounter, a region within the gravitational radius of the intruder is impacted
(Figure \ref{fig7}b).
As the intruder loops around the massive star, the disk develops transient spiral
arms where gas is strongly shocked (Figures \ref{fig7}c$-$e).  Parts of the outer
disk are ejected and the intruder steals material to 
form its own disk with an outer radius roughly given by its gravitational radius
near periastron (Figures \ref{fig7}e and f).  Figure \ref{fig7}g to i show three frames 
of this encounter from the vantage point of an observer located in the plane of 
the initial disk around  the massive star.    The disks relax within a few orbit times
at the disk outer radii.   

The Src I and BN disks have spin axes roughly at right-angles to their proper 
motions.    Within a the gravitational radius,  abrupt acceleration of a star will  
re-orient bound orbits.    
In the center of mass  frame of a recently accelerated star, disk particles  
experience a torque which tends to align their orbit planes so that after relaxation, 
the disk angular  momentum vector will tend to be at right-angles to the stellar 
proper motion  vector.  

\begin{figure*}
\gridline{\fig{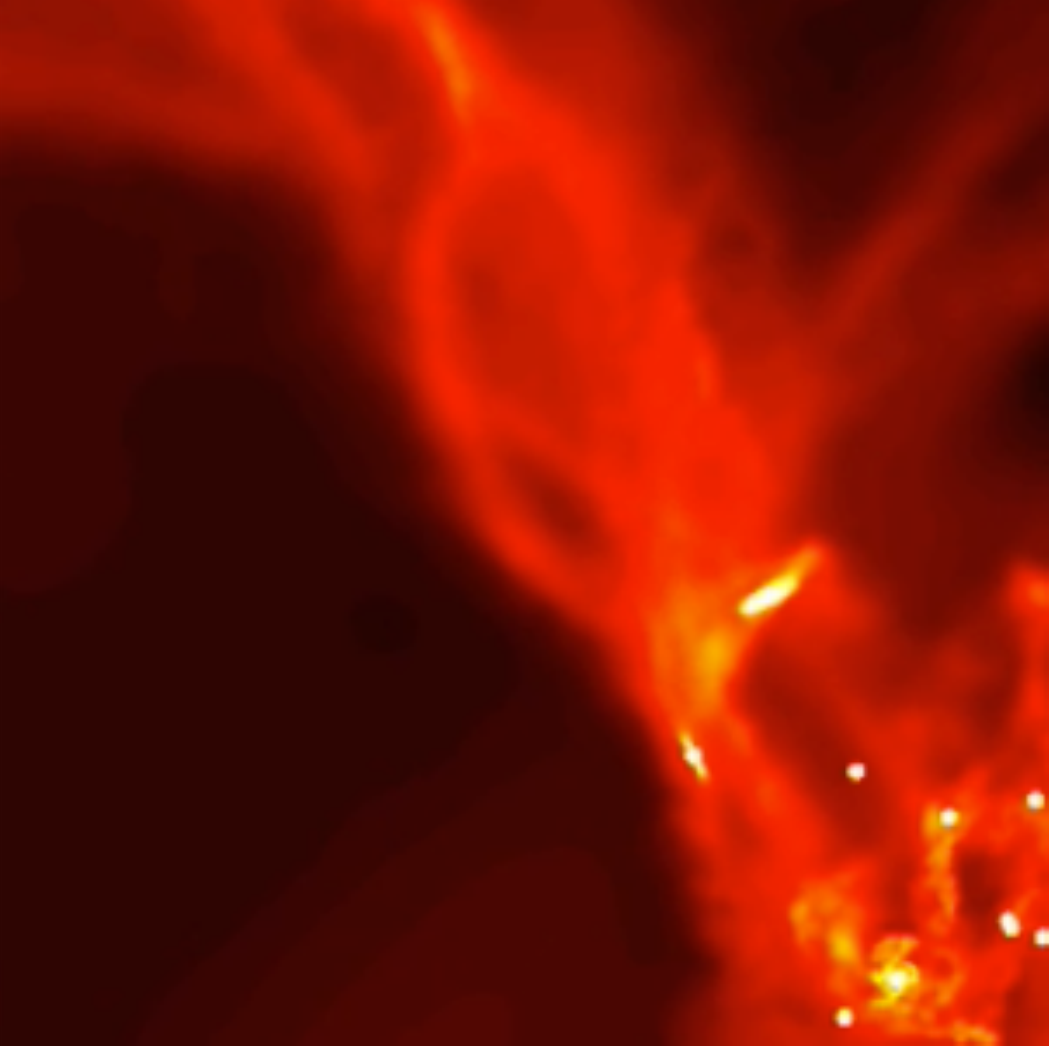}{0.33\textwidth}{(a)}
              \fig{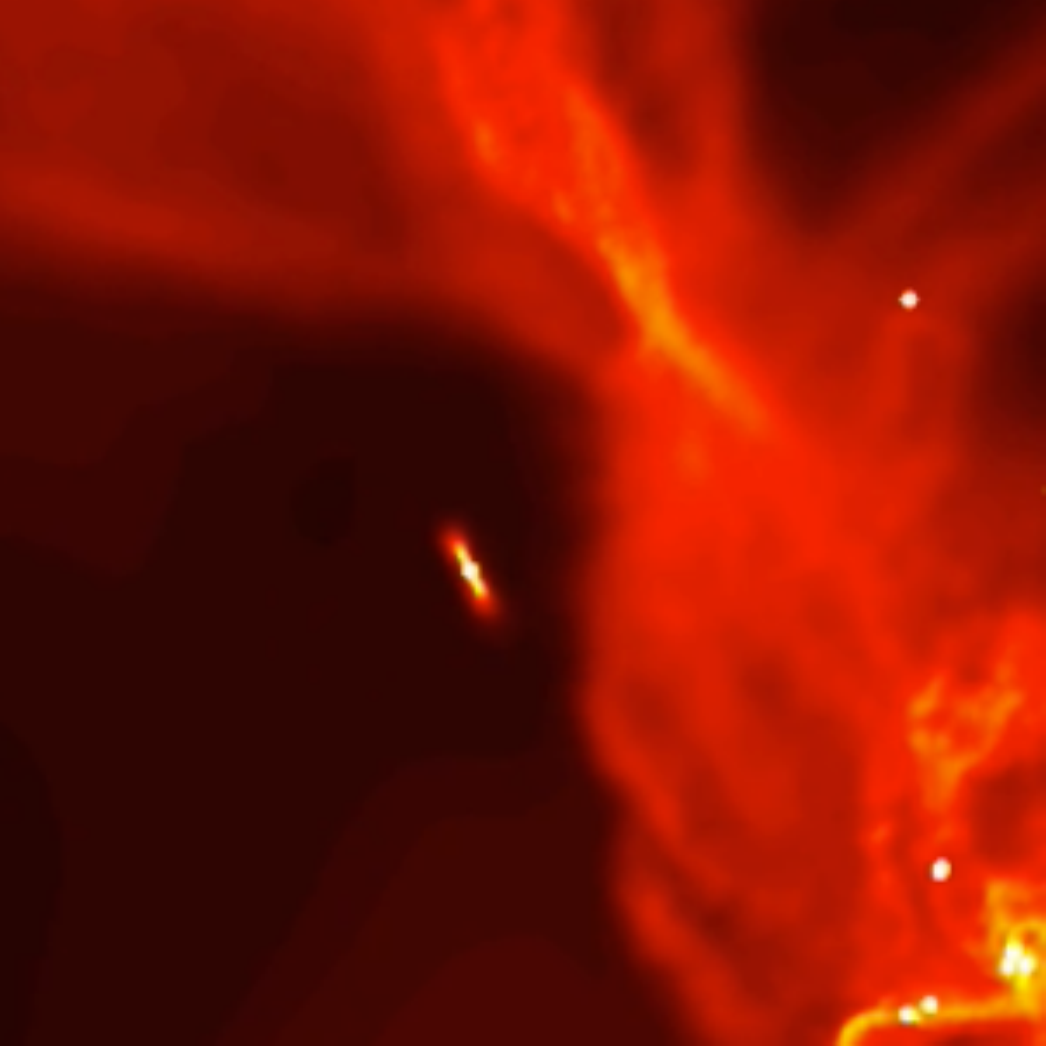}{0.33\textwidth}{(b)}
              \fig{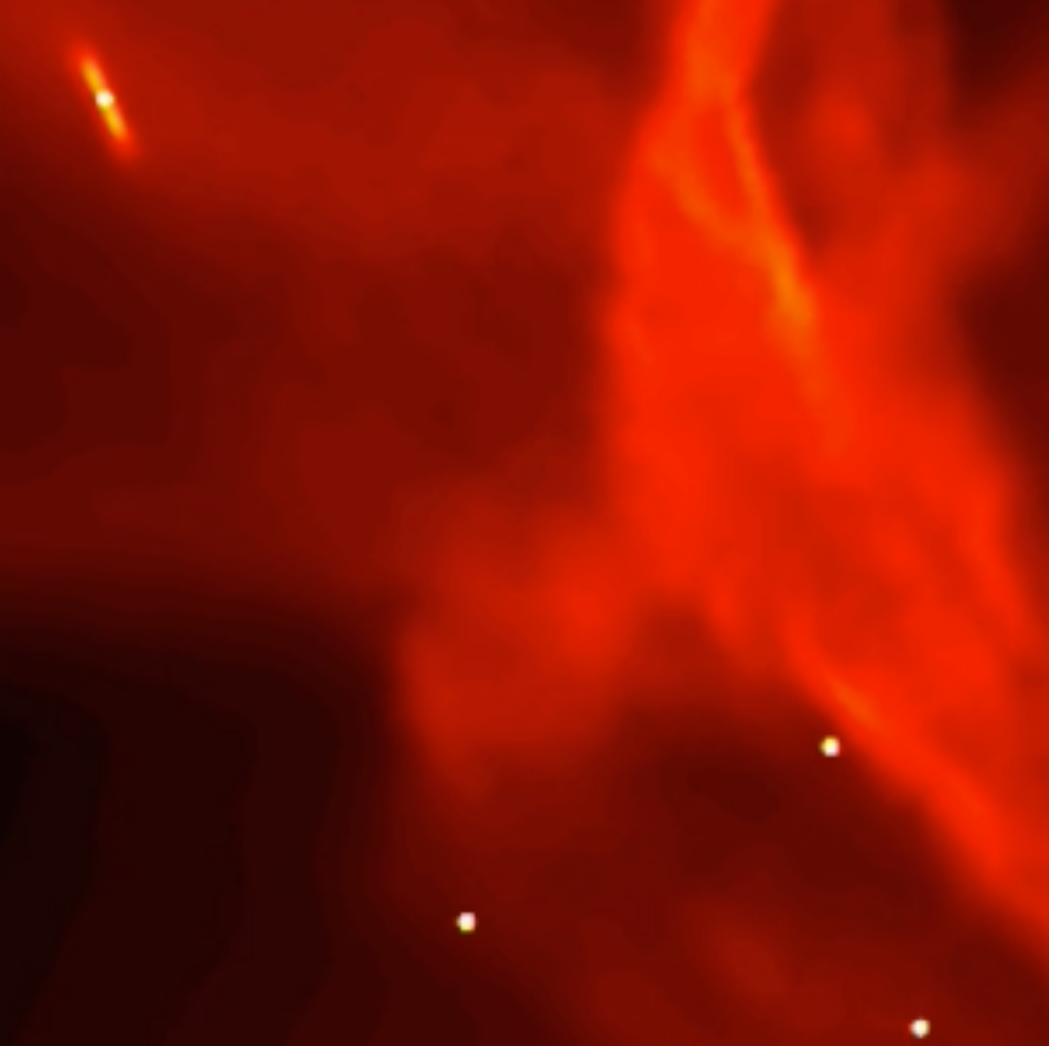}{0.33\textwidth}{(c)}
              }
\caption{Three frames from an sph simulation showing the final re-orientation of a disk around an ejected star.  Taken from \citep{Bate2002a,Bate2002b}.
\label{fig8}}
\end{figure*}

Figure \ref{fig8} shows three panels from the 3D simulations
described by \citet{Bate2002a,Bate2002b}.    A dynamically ejected star moves
from a cluster located in the lower-right of the figure to the upper left.   The 
original disk is re-oriented by the acceleration so that its spin-axis is nearly
orthogonal to the proper motion vector.   

The expected relationship between disk orientation and proper motion can
identify stars ejected by a dynamical decay of a multiple system.    
Disk major axes pointing  to a common  location may indicate dynamical decay.   
Lower mass stars can be ejected  with speeds of only a few kilometers per second 
\citep{ReipurthMikkola2010,ReipurthMikkola2012,ReipurthMikkola2015}.  
Disk orientations and proper motions can be measured with mm and sub-mm 
interferometers such as ALMA, NOEMA, or SMA, or using visual and infrared 
scattered light measurements \citep{Jiang2005} combined with astrometry.

\section{Summary and Conclusions}

The main results from this study are:

1) Inspection of multiple archival infrared images confirm the high proper motion of 
the moderate mass young star, Src~x, away from the point of origin of the
OMC1 explosion.   These data show that in the infrared, Src n does not
exhibit any measurable motion, confirming the proper motion detected at
radio wavelengths likely traces a small-scale one-sided ejection of plasma.
Although saturated in the IR images, BN also can be seen to move towards the 
northwest as shown by much more precise radio measurements. 

2) Analysis of the proper motion of the ejected stars BN, Src~I,
and Src~x, combined  with the requirement
that the center of mass of the three stars move with a speed less than
the escape speed from OMC1 shows that  Src I is likely to have a mass 
around 15 \Msol , consistent
with recent measurements of its disk rotation curve \citep{Ginsburg2018}.

3) The measured radial velocities of Src~I and BN, combined with the requirement
that the center of mass of the three ejected stars move with a speed less than
the escape speed from OMC1 indicates that Src~x is likely to have a radial 
velocity of $V_{LSR} \approx -23\pm 10$ \kms  .

4) The sizes, masses, and orientations of the Src I and BN disks, and limits on the 
Src~x disk, are consistent with the dynamical ejection scenario.   Matter
orbiting the accelerated star with a Kepler speed larger than the stellar 
ejection speed will tend to be dragged along.    Thus, the
current disks are likely to be inner parts of pre-existing disks that survived
the dynamic ejection $\sim$550 years ago. 

5) The high proper motion, compact radio sources dubbed `IRc23'  and 
Zapata 11  by \citet{Dzib2017} and called E2 and SW1 here do not contain 
compact the X-ray, IR,  or sub-mm sources at the sensitivity limit of existing data.  
Thus, they likely contribute negligible mass and momentum to the motion of 
the center-of-mass analysis.

6)  These slightly extended,  free-free emitting radio knots are part of a chain
of fainter radio sources which trace a younger than 500 year-old, collimated
outflow from BN with a mass-loss rate of $\dot M > 10^{-7}$ \Msol \yr .  As this 
ejecta moves through dense molecular gas
located behind the Orion Nebula, it is  likely to be collisionally ionized 
by shocks with speeds of  30 to 50  \kms , consistent with proper motion 
measurements.

\acknowledgments
This work was supported in part by National Science Foundation (NSF) grants AST-1009847, AST-1313188, and AST-1910393.  This paper makes use of the following ALMA data: ADS/JAO.ALMA $\#$2013.1.00546.S.    ALMA is a partnership of ESO (representing its member states), NSF (USA) and NINS (Japan), together with NRC (Canada), MOST and ASIAA (Taiwan), and KASI (Republic of Korea), in cooperation with the Republic of Chile. The Joint ALMA Observatory is operated by ESO, AUI/NRAO and NAOJ. The National Radio Astronomy Observatory is a facility of the National Science Foundation operated under cooperative agreement by Associated Universities, Inc.  
We thank Professor Norio Kaifu from providing a FITS copy of the Subaru \Htwo\ image from \citep{Kaifu2000}.  Some of the data use in the analysis is based on observations obtained with the Apache Point Observatory 3.5-meter telescope, which is owned and operated by the Astrophysical Research Consortium.  We thank the referee, Dr. Jonathan Tan  for helpful comments which improved the manuscript.

\vspace{5mm}
\facilities{Apache Point Observatory, Subaru, Gemini South, ALMA, JVLA}

\software{CASA \citep{McMullin2007}, IRAF \citep{Tody1986,Tody1993},  PyRAF (Science Software Branch at STScI 2012),   SAO Image ds9 \citep{JoyeMandel2003}
}


\begin{deluxetable*}{lllll}[b!]
\tablecaption{Summary of Near-IR Data Used in this Study}
\tablecolumns{5}
\tablenum{1}
\tablewidth{0pt}
\tablehead{
 \colhead{Telescope} &
 \colhead{Date} &
 \colhead{MJD} &  
 \colhead{Wavelength} & 
 \colhead{Reference}
}
\startdata
AAT			&  15 Aug 1992		& 48849		   & 1.644, 2.12 $\mu$m	&  \citep{AllenBurton1993} \\
Subaru		&  11-13 Jan 1999 	& $\sim$51190    & 2.12 $\mu$m  		&  \citep{Kaifu2000}   \\
APO 3.5 m	&   22 Nov 2004	& 53331	    	    & 2.12 $\mu$m		& \citep{Bally2011} \\
Gemini-S		&  25 Jan 2004		& 53029   		    & 11.7  $\mu$m			& \citep{Smith2005} \\
Gemini-S		&  30 Dec 2012 - 28 Feb 2013 & $\sim$56321  & 1.644, 2.12 $\mu$m	&  \citep{Bally2015} \\
\enddata
\label{table1}
\end{deluxetable*}
       

\begin{deluxetable*}{llllllll}[b!]
\tablecaption{Positions and proper motions of the runaway stars in OMC1}
\tablecolumns{7}
\tablenum{2}
\tablewidth{0pt}
\tablehead{
\colhead{Source} & 
\colhead{R.A.} & 
\colhead{Dec.} &
\colhead{$\mu (\alpha)$ cos($\delta$)} &
\colhead{$\mu (\delta)$} & 
\colhead{V}  & 
\colhead{P.A.} &
\colhead{Comments}
\\
\colhead{} &
\colhead{J2000} &
\colhead{J2000} &
\colhead{mas}   &
\colhead{mas }   &
\colhead{\kms }   &
\colhead{ \arcdeg } &
\colhead{ }
}
\startdata
BN 			&  5:35:14.110 	&  $-$5:22:22.70 	&  -8.1$\pm$0.4   & 10.8$\pm$0.4  & 25.6$\pm$0.8  &  323$\pm$2 		& \citep{Rodriguez2017} \\
			&			&				&  -6.9$\pm$1.4   & 10.3$\pm$1.4  & 23.5$\pm$4 	& 326$\pm$10		& \citep{Luhman2017} \\
			&			&				&  -4.3$\pm$1.1   & 12.7$\pm$1.1  & 25.4$\pm$2.1 	& 341$\pm$5	 	& \citep{Goddi2011} \\
			&     			&				&  -4.6$\pm$1.1   & 11.2$\pm$1.1  & 23.0$\pm$2.1   & 338$\pm$5            & Transformed to Dzib ONC frame \\
			&			&				&  -7.2$\pm$2.7   & 12.2$\pm$1.9  &	 26.9$\pm$3.3  	& 330$\pm$8		& \citep{Kim2018} \\
			&                       &                               &  -7.95$\pm$2    &  9.45$\pm$2    & 23.4$\pm$5    	& 320$\pm$15		& this work \\ 
Src I			&  5:35:14.520  &  $-$5:22:30.65 	&  1.8$\pm$0.4    &  -4.6$\pm$0.4  & 9.4$\pm$0.8 	& 159$\pm$5  		& \citep{Rodriguez2017} \\	
			&			&				&  5.5$\pm$1.1    &  -1.9$\pm$1.1  & 11.0$\pm$2.3  	& 109$\pm$12 		& \citep{Goddi2011} \\
			&			&				&  5.2$\pm$1.1    &  -3.4$\pm$1.1  & 11.8$\pm$2.3  	& 123$\pm$12 		& Transformed to Dzib ONC frame \\ 
Src x 		& 5:35:15.215  &   $-$5:22:37.04 	&  23.3$\pm$1.4  &  -17.4$\pm$1.4 & 55.1$\pm$4 	& 127$\pm$3		& \citep{Luhman2017} \\
			&			&				&  26.8$\pm$1.5  & -18.4$\pm$1.5  & 61.6$\pm$5	&  125$\pm$4		& \citep{Kim2018} \\
			&			&				&  20.2$\pm$3     &  -12.0$\pm$3   & 44.6$\pm$8 	& 120$\pm$6 		& this work	\\
Src n  		& 5:35:14.361	& $-$5:22:32.75	&  0.0$\pm$3       &  0.0$\pm$3      & 0.0$\pm$8 	&   	- 			& this work  \\
			&			&				&  1.9$\pm$1.0    &  1.0$\pm$0.7  &	4.0$\pm$3	& 62$\pm$40 		& \citep{Kim2018} \\
			&			&				&  -1.8$\pm$1.4   & -2.5$\pm$1.4   & 5.8$\pm$3	& 215$\pm$40 		& \citep{Luhman2017} \\
			&			&				&   $<$3          &   $<$3                 & $<$8			& -             		& \citep{Goddi2011} \\
			&			&				&   0.0$\pm$0.9   & -7.8$\pm$0.4   & 14.8$\pm$1.2	& 180.0$\pm$4		& jet from n \citep{Rodriguez2017} \\
SW1 (Zapata 11)  	& 05:35:14.019	& $-$5:22:23.25	&  -11.8$\pm$5.6 	& 9.3$\pm$2.3   & 28.4$\pm$11    & 308$\pm$20		&  PM from \citep{Dzib2017} \\
SW1 		  	& 05:35:14.107	& $-$5:22:23.20	&  -22.1$\pm$9.0 	& 5.3$\pm$5.0   & 43.1$\pm$20    & 283$\pm$30		&  PM from this work\\
SW2 (152)		& 05:35:13.953	& $-$5:22:23.82	&  17.2$\pm$9.4   	& -8.8$\pm$5.8  & 36.6$\pm$21    & -117$\pm$30   		& PM from this work \\
SW3 (147)		& 05:35:13.916	& $-$5:22:23.93	&  -31.8$\pm$4.1      & 10.1$\pm$4.6 & 63.2$\pm$12    &  288$\pm$15   		& PM from this work \\
SW4 (6 cm blob)	& 05:35:13.751	& $-$5:22:25.81	&       -        	 	&       -                & - 			     &  -   				&                               \\
E1				& 05:35:14.532	& $-$5:22:20.71	& -141.1$\pm$48 	&  96.5$\pm$35   & 324$\pm$113  &  304$\pm$30         	&  PM from this work \\	
E2 (`IRc23')		& 05:35:14.609 & $-$5:22:21.02	&  3.36$\pm$0.4  	&  20.1$\pm$8.6  & 38.7$\pm$16  & 9$\pm$15			&  PM from \citep{Dzib2017} \\
E2 				& 05:35:14.612 & $-$5:22:20.91	&  12.62$\pm$2.7  	&  21.2$\pm$3.7  & 46.7$\pm$9   & 31$\pm$20  		&  PM from this work  \\
\enddata
\tablecomments{All motions are in the Orion reference frame measured with respect to the mean proper motion (PM) of
the radio positions of low-PM sources measured with the VLA in C-band at 4 to 8 GHz.  
The absolute PM of the Orion reference frame moves with 
$\mu _{\alpha}$ cos $\delta$ = 1.07$\pm$0.09 mas~yr$^{-1}$, $\mu _{\delta}$ = -0.84$\pm$0.16~mas~yr$^{-1}$
\citep{Dzib2017}.   The absolute proper motions reported by \citep{Goddi2011} have been transformed to the \citep{Dzib2017}
OMC1 reference frame.  A distance of D = 400 pc is assumed for the conversion of proper motions in the OMC1 frame to
velocity. }
\label{table2}
\end{deluxetable*}

\begin{deluxetable*}{lrrrc}[b!]
\tablecaption{Parameters for the Ejected Stars in the OMC1 Reference Frame}
\tablecolumns{5}
\tablenum{3}
\tablewidth{0pt}
\tablehead{
 \colhead{Quantity } &
 \colhead{Src I} &
 \colhead{BN} &  
 \colhead{Src x} & 
 \colhead{Unit}
}
\startdata
V$_{PM}$		&   9.4	& 25.6	& 55.1	&   \kms  \\
M			&  15		& 12		&   3		&   \Msol  \\
V$_r$		& $-$4 	& 14.2	& $-$32	&   \kms \\
V$_{tot}$		& 10.2	& 29.3	& 60		&   \kms \\
K.E.			&  1.6	& 7.8		& 91		& $\times 10^{46}$ erg \\
\enddata
\label{table3}
\end{deluxetable*}

\begin{deluxetable*}{lrrccl}[t!]
\tablecaption{Proper Motions$^a$ of suspected BN Outflow Features Relative to BN.}
\tablecolumns{6}
\tablenum{4}
\tablewidth{0pt}
\tablehead{
   \colhead{Object}                    & 
   \colhead{{$\mu_\alpha \cos(\delta)$}} & 
   \colhead{{$\mu_\delta$}}        & 
   \colhead{$V_t$}                    & 
   \colhead{P.A.}                       & 
   \colhead{Comments}
   \\
   \colhead{ }                             &  
   \colhead{{$(mas\ yr^{-1})$} }   & 
   \colhead{{$(mas\ yr^{-1})$}}    & 
   \colhead{{$(km\ s^{-1})$} }      & 
   \colhead{$(\deg)$ }                  &  
   \colhead{} 
 }
 \startdata
 \hline
SW1     &  $-14.0\pm9.4$         & $-5.4\pm5.4$          & $28.4$   & $248$     & Close to BN\\
SW2     &  $ 25.3\pm9.8$         & $-19.6\pm6.2$        & $60.7$   & $128$      & New component in 2016? \\
SW3     &  $-23.7\pm4.5$         & $-0.7\pm5.0$          & $44.9$   & $268$      & Brightened in 2016 \\
E1$^b$ &  $-133\pm50$           & $85.7\pm35$          & $300 $    & $303$     & Bow with photometric variations \\
E2        &  $ 20.7\pm3.0$         & $10.4\pm4.0$         & $43.9$    & $63$        & Bright knot near IRc23 \\
\enddata
  \tablenotetext{a}{Proper motions measurements from Vargas-Gonz\'{a}lez et al. (in preparation) relative to the proper motion of BN determined by \citet{Rodriguez2017}.   The proper motion of BN is assumed $\mu_\alpha \cos(\delta)$=$-$8.1 mas~yr$^{-1}$,  $\mu_\delta$=+10.8 mas~yr$^{-1}$. }
  \tablenotetext{b}{Source not considered in the catalogs of  Forbrich et al. (2016)  or Vargas-Gonz\'{a}lez et al. in prep. ($SNR<5$).}
  \label{table4}
\end{deluxetable*}

\begin{deluxetable*}{lllcccl}[b!]
\tablecaption{Parameters of the BN Radio Jet}
\tablecolumns{7}
\tablenum{5}
\tablewidth{0pt}
\tablehead{
 \colhead{Object} 			&
 \colhead{Peak S$_{\nu}$} 	& 
 \colhead{Total S$_{\nu}$} 	&
 \colhead{Size} 				&
 \colhead{D$_{BN}$}  		&
 \colhead{n$_e$} 			& 
 \colhead{Comments}
 \\
 \colhead{} 				&
 \colhead{($\mu$Jy)} 		& 
 \colhead{($\mu$Jy)} 		&
 \colhead{(\arcsec )} 			&
 \colhead{(\arcsec )} 			&
 \colhead{ ($\times 10^4$ cm$^{-3}$)}  &
 \colhead{}
}
\startdata
SW1 	&   270	&  416  	&  0.4\arcsec\ $\times$  0.8\arcsec\  	& 1.5		& 5.2		& 157$^a$ (Zapata 11)$^b$ \\
SW2 	&   64	&     41 	&  0.3\arcsec\   					& 2.7		& 1.9		& 152$^a$ \\
SW3 	&   80	&     50 	&  0.3\arcsec\  	 				& 3.1		& 2.1		& 147$^a$ \\
SW4		&   25	&   260 	&  1.6 \arcsec\ 	 				& 6.3		& 2.0		& 6 cm blob \\
BN   		&   2320	&  2473 	&  0.6\arcsec\   					&  - 		&  -		&  \\
E1		&   78	&  2650	&  0.3\arcsec\ $\times$  0.67\arcsec\  & 6.9	& 15		& bow near IRc23	 \\
E2	 	&   47	&  1048 	&  0.25\arcsec\ $\times$  0.40 \arcsec\  & 7.7	& 10		&  `IRc23' $^b$  \\
\enddata
 \tablenotetext{a}{Designation in \citep{Forbrich2016}}
 \tablenotetext{b}{Designation in \citep{Dzib2017}}
 \label{table5}
\end{deluxetable*}

\newpage

\end{document}